\documentclass[]{aa}
\usepackage{graphicx}
\usepackage{times}
\begin{document}
\newcommand{\bull}{\vrule height .9 ex width .8ex depth -.1ex \hskip 0.2ex}
\newcommand{\blok}{$\clubsuit$ $\clubsuit$ $\clubsuit$}
\thesaurus{03(11.06.2; 11.16.1; 11.19.2)}
\title{Surface photometry of bulge dominated low surface brightness galaxies}
\author{M. Beijersbergen\inst{1}  \and W.J.G. de Blok\inst{2} \and J.M. van der Hulst\inst{1}}
\institute{Kapteyn Astronomical Institute, P.O. Box 800, 9700 AV
       Groningen, 
       The Netherlands \and ATNF, P.O. Box 76,
       Epping, NSW 1710, Australia}
\offprints{M. Beijersbergen} 
\date{Received date; accepted date}

\maketitle

\begin{abstract}

We present results of broad band \emph{BVRI} observations of a sample of
galaxies with a low surface brightness (LSB) disk and a bulge. These galaxies are well
described as exponential disks and exponential bulges with no
preferred value for either scale length or central surface
brightness. The median \emph{B} band disk scale length is
12.6 kpc ($H_{0}$ = 75 km s$^{-1}$ Mpc$^{-1}$) 
which is much larger than scale lengths of typical (disk dominated) LSB or high
surface brightness (HSB) galaxies. Furthermore, the disk and bulge
scale lengths are correlated, suggesting a coupling in the formation. Bulge dominated LSB galaxies are observed to be redder than 
disk dominated LSB galaxies and
their bulge-to-disk ratios are increasing towards redder
wavelengths. We find colors that are comparable to or bluer than HSB galaxies
of the same morphological types. Bulge dominated LSB galaxies are therefore not faded
HSB galaxies with no current star formation. We find that bulge dominated LSB galaxies fit in with
the general trends defined by the HSB galaxies. 
The properties of these bulge dominated LSB galaxies show that LSB
galaxies do not just come in two varieties. They cover the entire
range in optical and morphological properties between late-type disk
dominated LSBs and giant Malin-1-like LSBs. LSB galaxies thus also
form a LSB Hubble sequence, parallel to the classical HSB one.

\keywords{Galaxies: fundamental parameters -- Galaxies: photometry --
Galaxies: spiral}
\end{abstract}

\section{Introduction}

When studying the properties and evolution of galaxies it is necessary
to have  a proper census of all types of galaxies. Deep photographic
surveys have shown the existence of a large number of galaxies with
surface brightnesses much fainter than the night sky. The severe
selection effects caused by the brightness of the night sky ensures
that these galaxies are very much under-represented in conventional
galaxy catalogs. However, in the last 20 years it has become clear
that these low surface brightness (LSB) galaxies may constitute a
major fraction of the total galaxy population. Clearly, LSB galaxies
show us an alternative path of galaxy evolution which does not lead to
the classical Hubble sequence and offer us a new window onto galaxy
evolution (Impey \& Bothun \cite{impey}; Bothun et al. \cite{bothun}).

Most of the LSB galaxies investigated in any detail are either
  late-type and disk dominated (de Blok et al. \cite{de blok et al},
  hereafter dB95;
  McGaugh \& Bothun \cite{mc gaugh}), or giant, Malin-1-like galaxies
  (Sprayberry et al. \cite{sprayberry}; Pickering et
  al. \cite{pickering}). The disk dominated LSB galaxies have only
  small traces of star formation (de Blok et al. \cite{db}a), are very
  gas-rich (de Blok et al. \cite{db2}b), and appear quite unevolved
  (van der Hulst et al. \cite{vdhulst}). The \ion{H}{i} masses are a few times
  $10^9 M_{\odot}$ (de Blok et al. \cite{db2}b) and the \ion{H}{i}
  surface densities are usually close to the critical density for star
  formation (van der Hulst et al. \cite{vdhulst}; de Blok et
  al. \cite{db2}b; Kennicutt \cite{kennicutt}). They also have color gradients, the outer parts of the disks are bluer than the inner
parts. Apart from this the LSB galaxies investigated by de
Blok, McGaugh and van der Hulst tend to be bluer than classical Hubble
sequence HSB galaxies. There has been some discussion on whether this
was an intrinsic property of LSB galaxies or yet another selection
effect, due to the blue-sensitive plates used for the LSB surveys. The
discovery of a large number of red LSB galaxies (O'Neil et
al. \cite{o'neil}) showed that there is also a red (\emph{B-V} $>$ 0.9) component of the LSB
population. The continuous range of colors from the very blue to the
very red clearly shows that LSB galaxies define a wide range of evolutionary
states. 

The fact that most LSB galaxies discovered so far are either
bulgeless, late-type ``normal-sized'' galaxies or giant galaxies with
a significant bulge component, raises the question whether there are
any ``normal-sized'', bulge dominated LSB galaxies. Are there LSB
galaxies with disks with scale lengths of a few kpc and a low surface
brightness that have a significant bulge component?

If these galaxies do indeed exist then the fact that they are not
turning up in LSB galaxy surveys can only mean that there are severe
selection effects against them, or that we already found them and
included them in our catalogs.

Because of their obvious bulge components it is hard to see which of the
known selection effects could make us very much biased against
them. We have therefore assumed that at least a fraction of the bulge
dominated LSB galaxies have already been included in conventional
catalogs, and have selected a small sample of what we can call bulge
dominated LSB galaxies from the ESO-LV catalog (Lauberts
\& Valentijn \cite{lauberts_valentijn}).

In the universe galaxies have surface brightnesses in
a continuous range from low to high surface brightness, the low end
being set by instrument sensitivity. Normally a LSB galaxy is \emph{defined} to be a galaxy with a face-on
\emph{disk} central
surface brightness more than one magnitude fainter than the Freeman
 (\cite{freeman}) value of $21.65 \pm 0.30$ \mbox{\emph{B} mag
arcsec$^{-2}$}.  In the presence of a bulge the true central surface
brightness of the galaxy will of course be much higher and it would be
hard to call them LSB for that reason. However, that is not the point
here. Bulge dominated galaxies can still have a LSB \emph{disk}, and
in that sense the central disk surface brightness is used as an
indicator for the amount of evolution in the disk. For late-type galaxies one can
roughly equate (for undisturbed galaxies) the central disk surface
brightness with evolutionary stage. The possibility of having a bulge
sitting in the middle of a LSB, possibly unevolved, disk raises
interesting questions as how can an evolved component such as a bulge
exist in an unevolved disk without affecting its evolution? Do bulges
and disks evolve independently in LSB galaxies? Are bulge and disk
surface brightnesses and sizes related? 
Galaxies towards the later types have a range in surface brightness
and size. The early-type galaxies are mostly HSB or giant LSB galaxies.
Do bulge dominated LSB galaxies form the ``missing
link'' between HSB and giant LSB galaxies?
 
To explore some of these questions we selected and observed a sample
of 20 LSB galaxies with bulges. This paper is organized in the
following way: in section 2 we discuss the
sample selection and reduction. The structural parameters, central surface brightness and scale length, will
be discussed in section 3.1 and magnitudes in section 3.2. The colors
and color profiles are described in section 3.3. These properties are
then discussed and compared to the properties
of disk dominated LSB and HSB galaxies in sections 4.1 and
4.2. A detailed look at the bulges is taken in section 4.3. We will summarize the discussion
and give concluding remarks in section 5. We define a LSB galaxy to be a galaxy with a face-on \emph{disk} central
surface brightness one magnitude fainter than the Freeman value. In
this paper we use $H_{0}$ = 75 km s$^{-1}$ Mpc$^{-1}$.

\section{The sample and reduction}

The sample of galaxies was taken from the ESO-LV catalog (Lauberts
\& Valentijn \cite{lauberts_valentijn}). Here we will briefly describe
the selection criteria with their justifications. 

The galaxies were selected to have types Sa--Im so that we are not biasing ourselves to
one person's/algorithm's classification scheme. The galactic latitudes
were chosen so as to diminish foreground extinction ($\mid b \mid$ $>
15\degr$) and the inclinations were selected to be smaller than
50$\degr$ to be not too sensitive to internal extinction. Furthermore, the surface brightnesses of the disks at
half light radius were selected to be fainter than 23.8 \emph{B} mag arcsec$^{-2}$
and the diameters of the 26 \emph{B} mag arcsec$^{-2}$ isophotes were selected to lie
between 1$\arcmin$ (pick large enough galaxies) and 3$\arcmin$
(chipsize limitation). These selection criteria
resulted in approximately 600 galaxies. All these galaxies were inspected and
selected on clean stellar foreground. From this subsample a random
sample of galaxies which had central light concentrations was chosen. The result of this selection is
that most of the galaxies in our sample have high surface brightness
bulges embedded in low surface brightness
disks. In the appendix \emph{R} band images of our sample of LSB galaxies are
shown. These images are presented using linear or exponential
intensity scale (see captions)  and
central parts may be saturated. It can be seen from these images that
we have roughly two types of bulge LSB galaxies in our sample. One
type has a normal, round bulge and the other type has bars and rings. In table~\ref{sample_galaxies} we list relevant sample
information. Column 1 contains the ESO-LV name of the galaxy, column 2 gives
the right ascension (1950.0) and column 3 the declination
(1950.0). Column 4 gives the distance (Mpc) as taken from the ESO-LV
catalog. Where not available, redshifts were determined with the
single dish Parkes telescope (de Blok et al. 1999, in preparation).  The inclination as derived from the data is
given in column 5 and total absolute \emph{B} magnitudes as derived from the
data are given in column 6. Hubble types are listed in column 7. 

All \emph{BVRI} images were taken with the 0.9 meter Dutch Telescope at La
Silla in October 1993, January 1994 and in March 1994. 
The CCD image reduction was done using standard procedures in the
Image Reduction and Analysis Facility (IRAF) and the Groningen Image
Processing SYstem (GIPSY; van der Hulst et al. \cite{vdhulst2}). The average seeing was $\sim
1\farcs5$ and the images have a limiting surface brightness
of $\sim26$ \emph{B} mag arcsec$^{-2}$. The main source of errors in
determining magnitudes and colors is the uncertainty in the sky level. We
measured the sky level in five boxes placed on parts of the image free
of stellar emission. The mean difference between the median sky levels
in these boxes was used as an estimate for the error introduced by
subtracting the sky. We determined the position angles and
inclinations from smoothed \emph{R} images. Stars and
cosmic ray defects were blanked prior to any fitting. We used fixed position angles, inclinations and
ellipse centers to make ellipse fits to the isophotes of the galaxies
and to integrate along these ellipses. Each band of a galaxy is thus
fit with the exact same model. All magnitudes and colors have been
corrected for atmospheric and galactic extinction. Extinctions in the
\emph{B} band were taken from the NED database.

\begin{table*}
\caption{Sample parameters.}
\begin{flushleft}
\begin{tabular}{lllllll}
\hline
(1)&(2)&(3)&(4)&(5)&(6)&(7)\\
Name&$\alpha(1950.0)$&$\delta(1950.0)$&D&i&$M_{\rm T,B}$&Type\\
\hline
2060140&06:27:10&-48:43:42&200&37&-21.91&SAc\\
1150280&02:42:31&-60:07:28&87&52&-20.57&SBc\\
1530170&02:03:21&-55:20:58&88&44&-20.86&SABbc\\
2520100&05:04:41&-45:06:45&133&41&-21.49&SBb\\
3500110&00:20:00&-34:23:52&200&26&-22.08&-\\
4220090&04:52:42&-28:44:12&63&24&-18.48&SBd\\
4250180&06:18:49&-27:53:36&91&33&-19.03&SABd\\
4990110&09:49:43&-25:04:36&37&34&-17.66&SBd\\
5450360&02:31:38&-21:15:44&231&31&-21.39&SAa\\
5520190&04:51:21&-18:00:12&163&52&-21.19&S?\\
1220040&06:41:38&-61:01:12&127&51&-20.50&SBa\\
0140040&02:42:13&-78:20:24&-&35&-&Sb\\
1590200&05:33:22&-55:54:48&-&27&-&SABd\\
3740090&09:52:49&-33:49:30&37&36&-17.80&Sa\\
4350310&09:59:16&-30:28:06&-&42&-&Sbc\\
0050050&07:33:35&-84:12:00&-&23&-&SABdm\\
0540240&03:51:32&-72:05:30&196&33&-20.77&SABab\\
5650160&09:31:54&-21:45:12&-&41&-&SB0/a\\
0590090&07:36:33&-70:35:48&20&28&-17.69&Sadm\\
4370420&10:39:08&-31:31:06&35&48&-18.43&SBc\\
4400490&12:02:59&-31:08:42&29&35&-&SABd\\
\hline       
\end{tabular}
\label{sample_galaxies}
\end{flushleft}
\end{table*}

\begin{figure}[!t]
\resizebox{\hsize}{!}{\includegraphics[height=5.6cm, width=8cm]{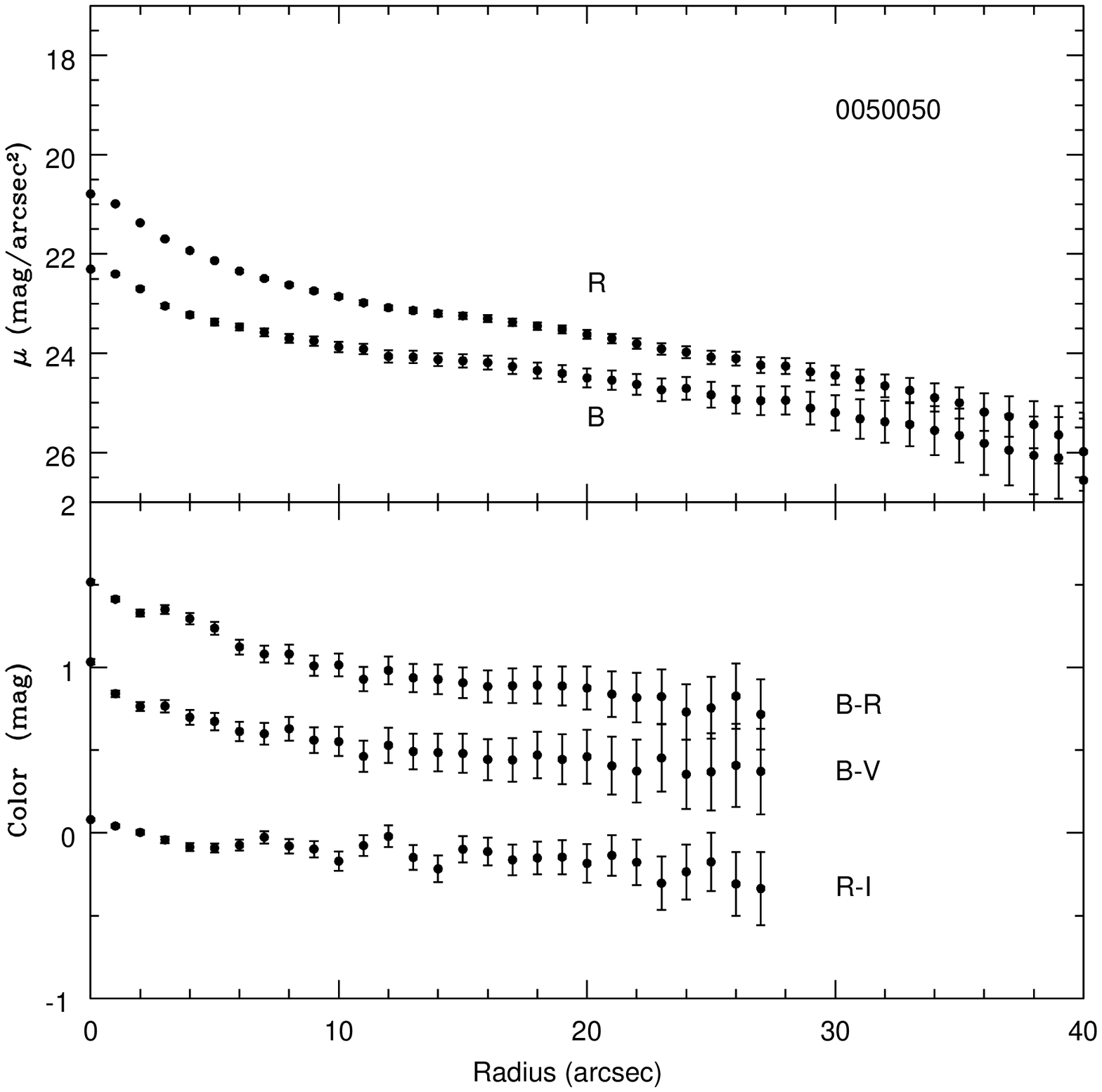}}
\resizebox{\hsize}{!}{\includegraphics[height=5.6cm, width=8cm]{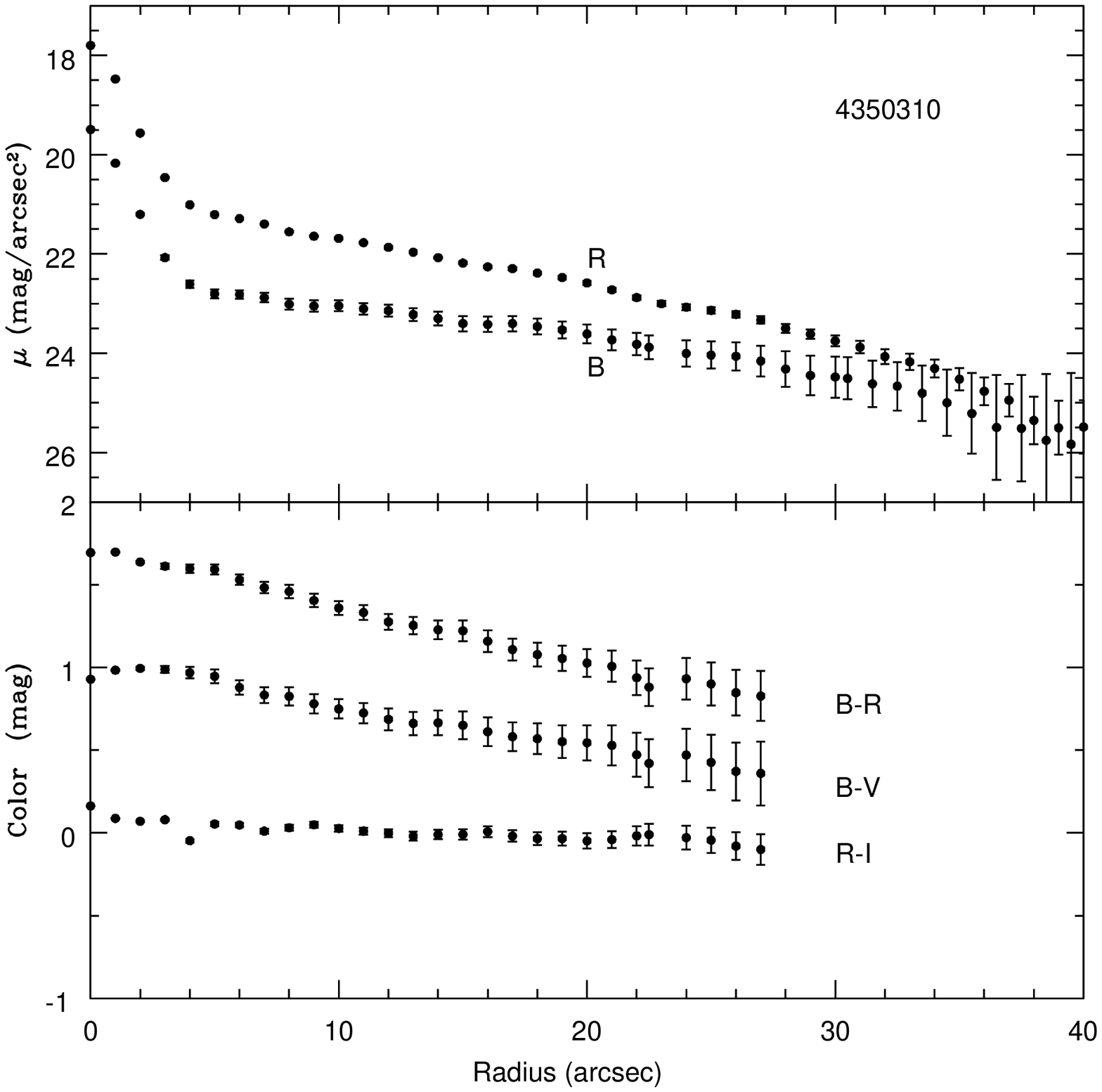}}
\resizebox{\hsize}{!}{\includegraphics[height=5.6cm, width=8cm]{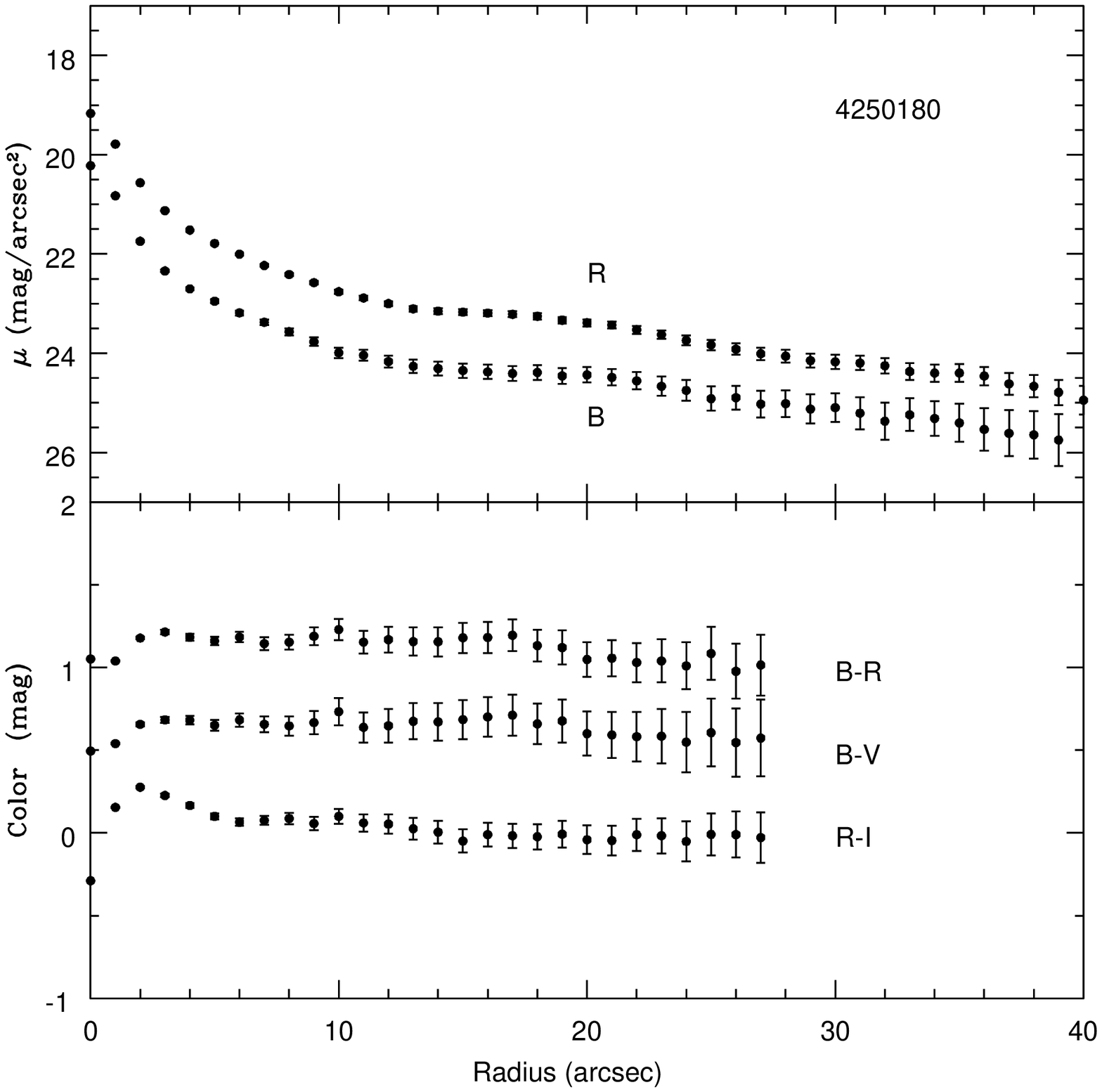}}
\caption[]{(a) Top panels contain \emph{B} and \emph{R} surface brightness profiles;
bottom panels contain radial color profiles for the galaxies
\object{ESO-LV 0050050}, \object{ESO-LV 4350310} and \object{ESO-LV
4250180}. In all cases \emph{R-I}
has been offset by -0.5 mag.}
\label{profiles}
\end{figure}
\begin{figure}[!t]
\setcounter{figure}{0}
\resizebox{\hsize}{!}{\includegraphics[height=5.6cm, width=8cm]{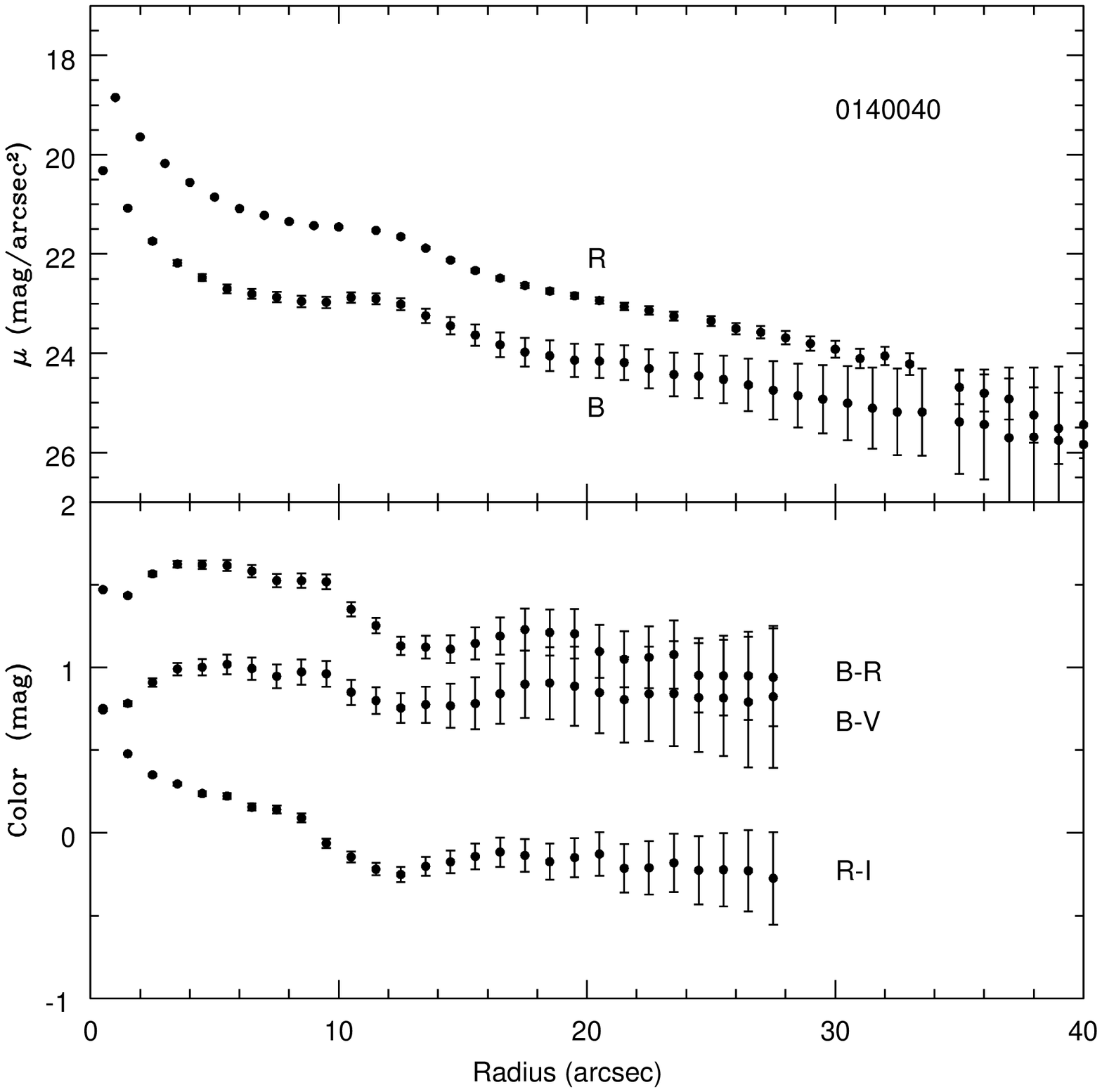}}
\resizebox{\hsize}{!}{\includegraphics[height=5.6cm, width=8cm]{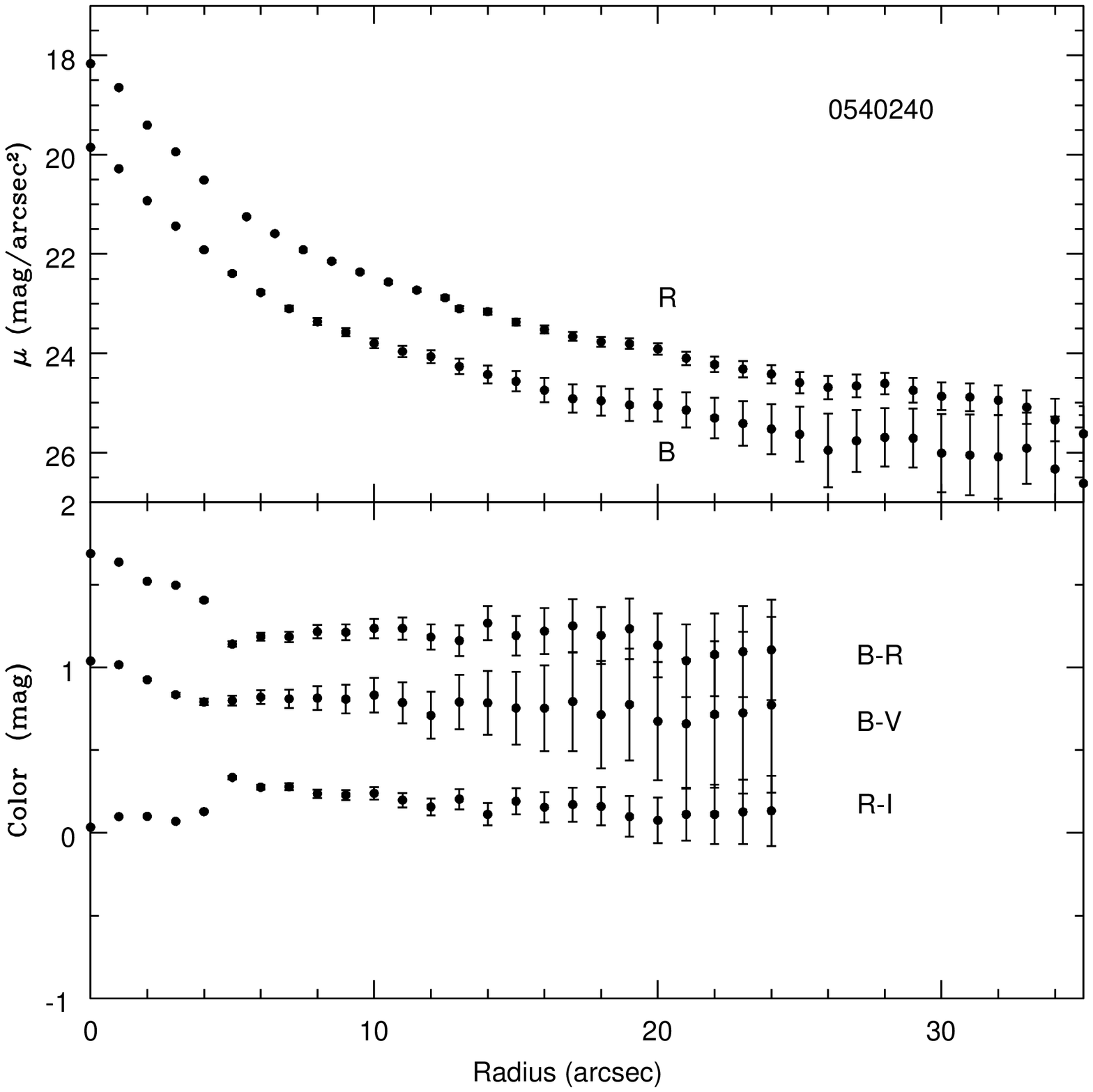}}
\resizebox{\hsize}{!}{\includegraphics[height=5.6cm, width=8cm]{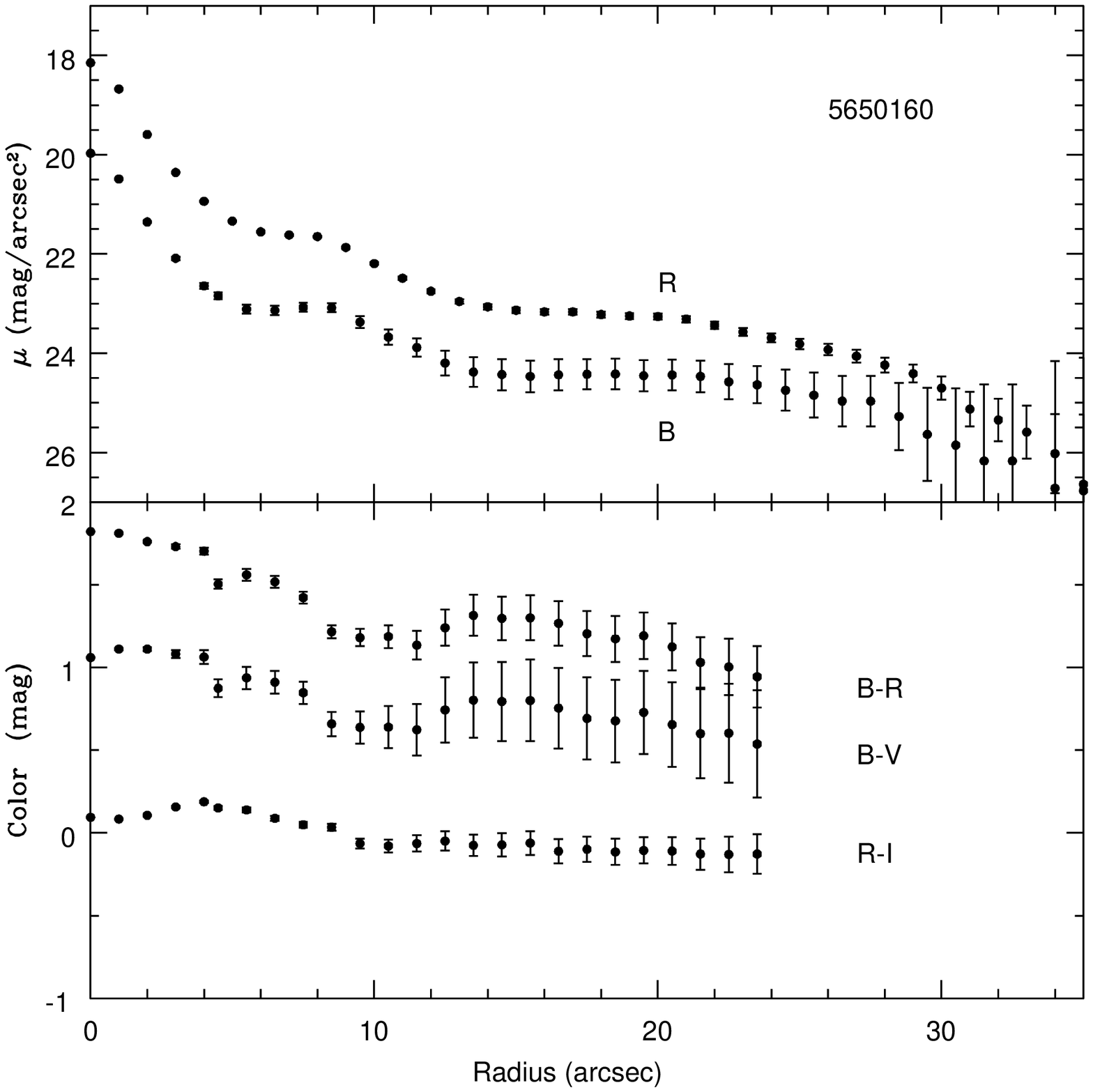}}
\caption[]{(b) Top panels contain \emph{B} and \emph{R} surface brightness profiles;
bottom panels contain radial color profiles of the galaxies
\object{ESO-LV 0140040}, \object{ESO-LV 0540240} and \object{ESO-LV
5650160}. In all cases \emph{R-I}
has been offset by -0.5 mag.}
\end{figure}
\begin{figure}[!t]
\setcounter{figure}{0}
\resizebox{\hsize}{!}{\includegraphics[height=5.6cm, width=8cm]{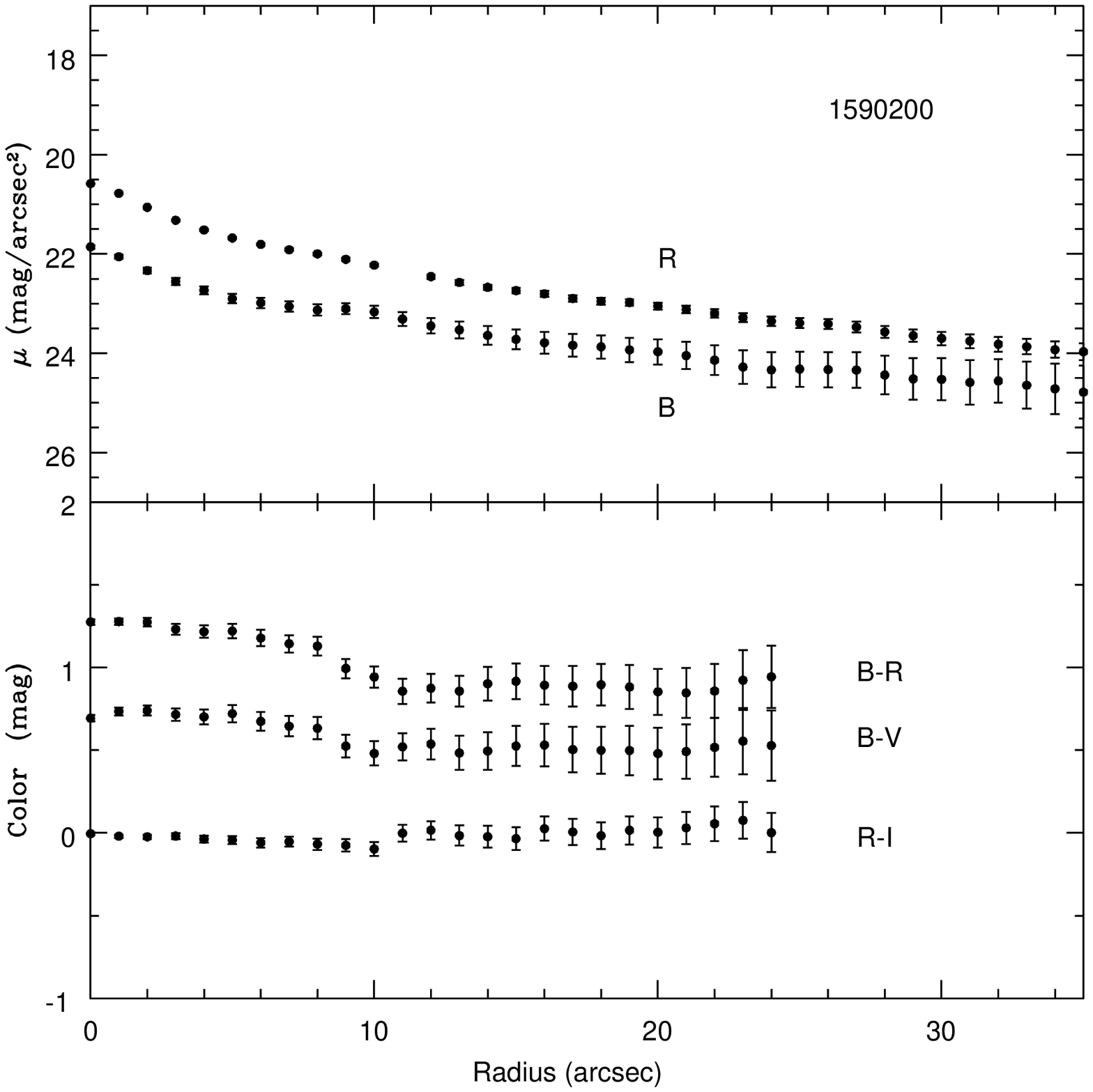}}
\resizebox{\hsize}{!}{\includegraphics[height=5.6cm, width=8cm]{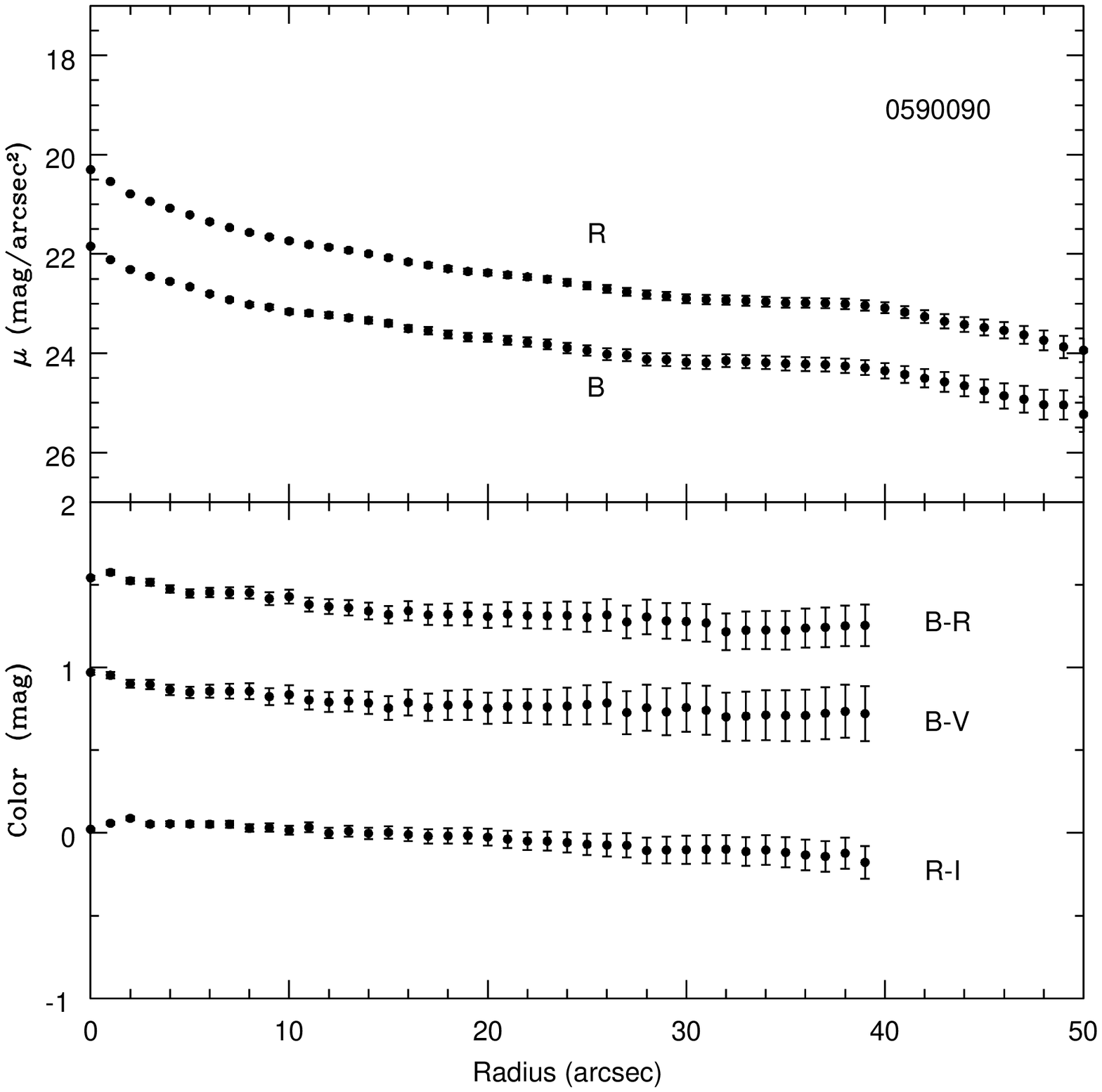}}
\resizebox{\hsize}{!}{\includegraphics[height=5.6cm, width=8cm]{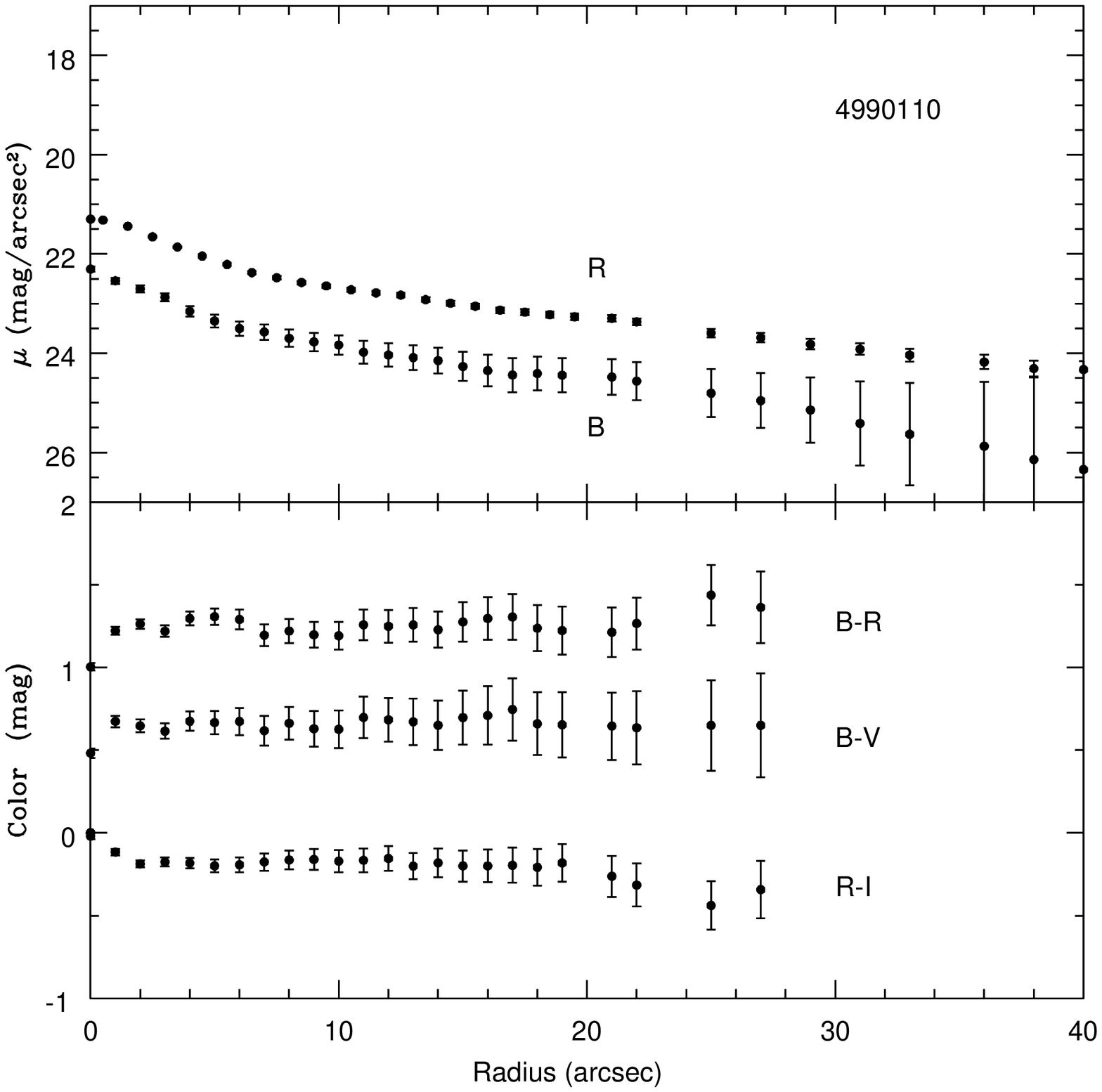}}
\caption[]{(c) Top panels contain \emph{B} and \emph{R} surface brightness profiles;
bottom panels contain radial color profiles for the galaxies
\object{ESO-LV 1590200}, \object{ESO-LV 0590090} and \object{ESO-LV
4990110}. In all cases \emph{R-I}
has been offset by -0.5 mag.}
\end{figure}
\begin{figure}[!t]
\setcounter{figure}{0}
\resizebox{\hsize}{!}{\includegraphics[height=5.6cm, width=8cm]{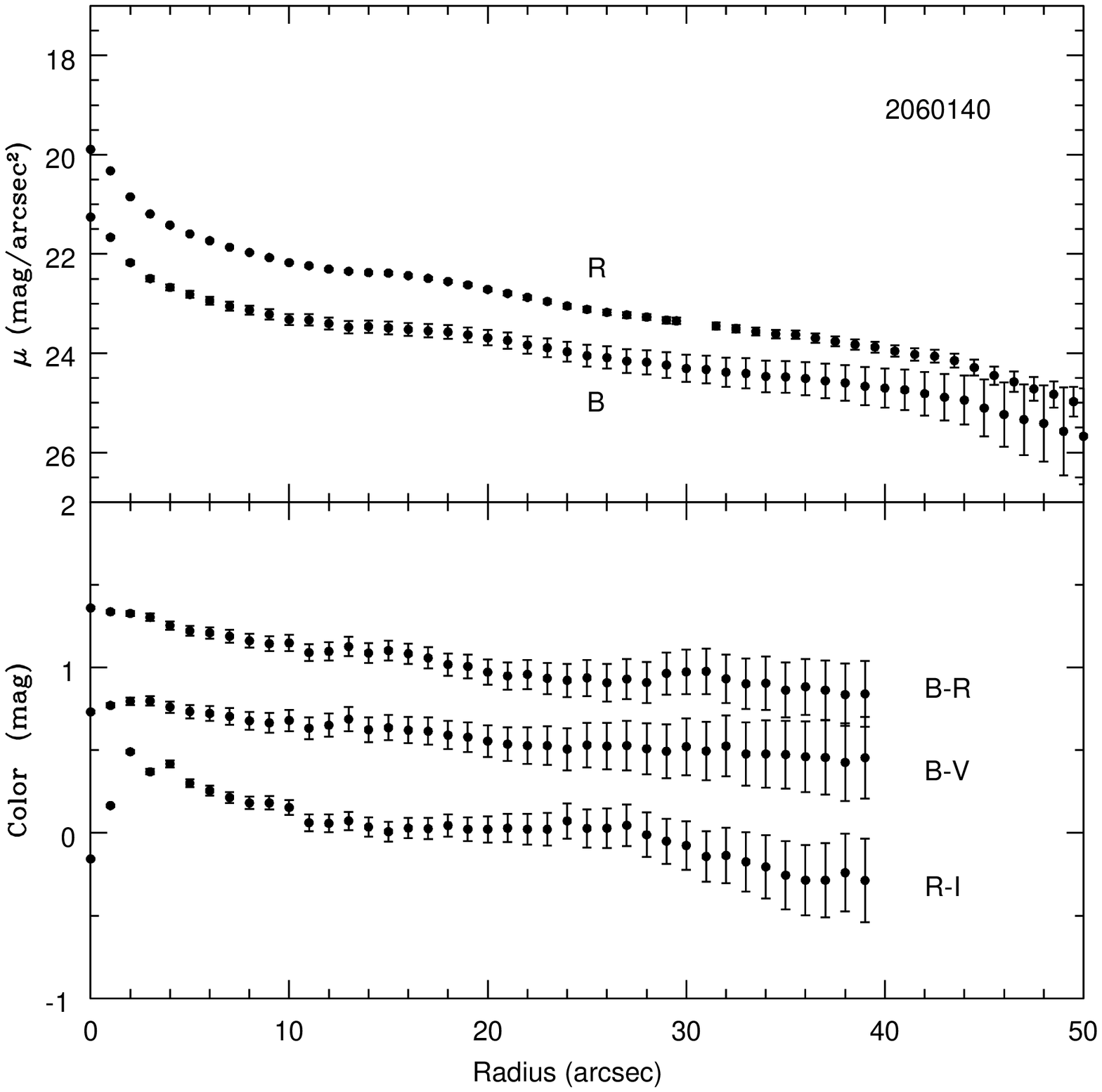}}
\resizebox{\hsize}{!}{\includegraphics[height=5.6cm, width=8cm]{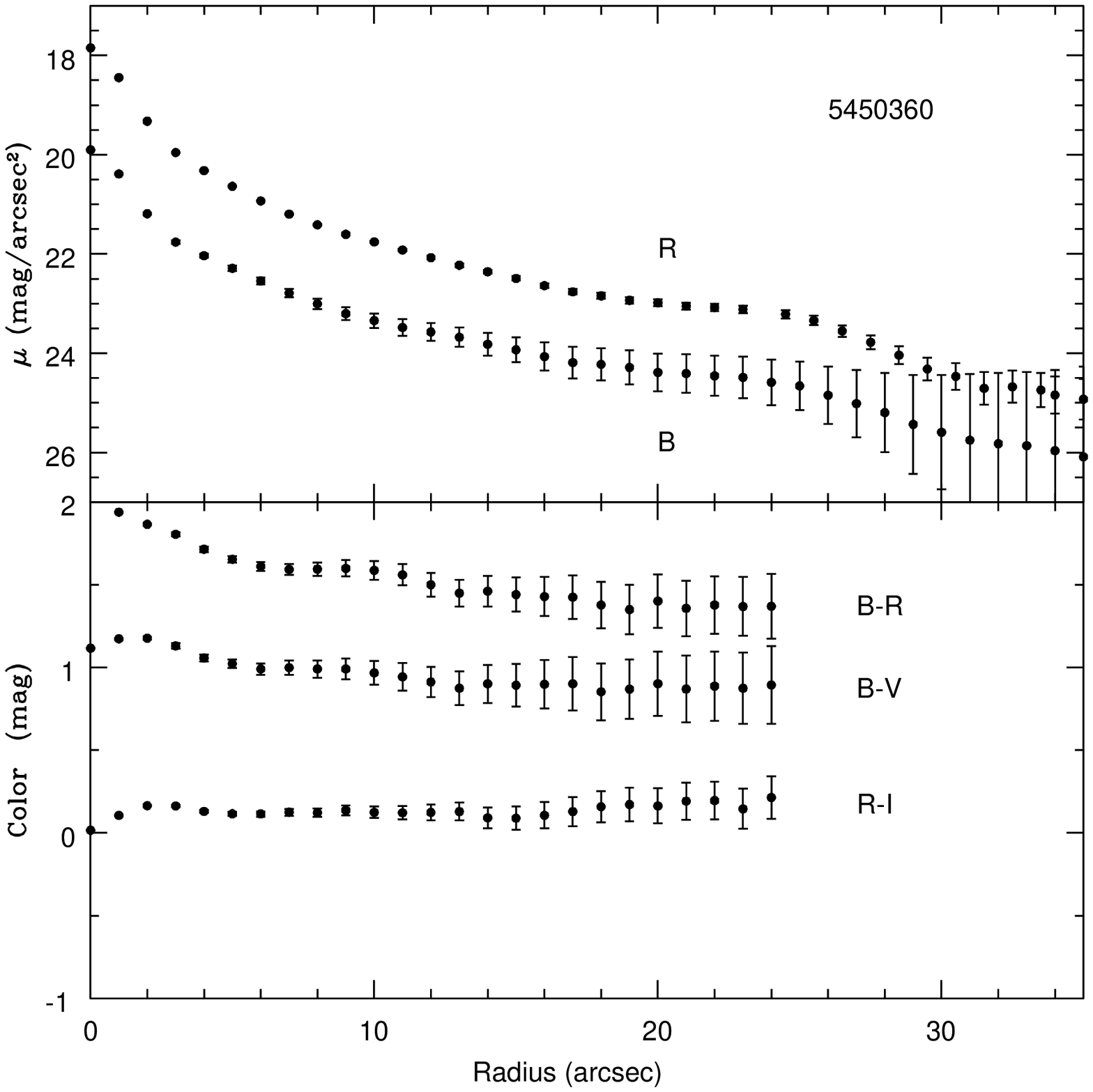}}
\resizebox{\hsize}{!}{\includegraphics[height=5.6cm, width=8cm]{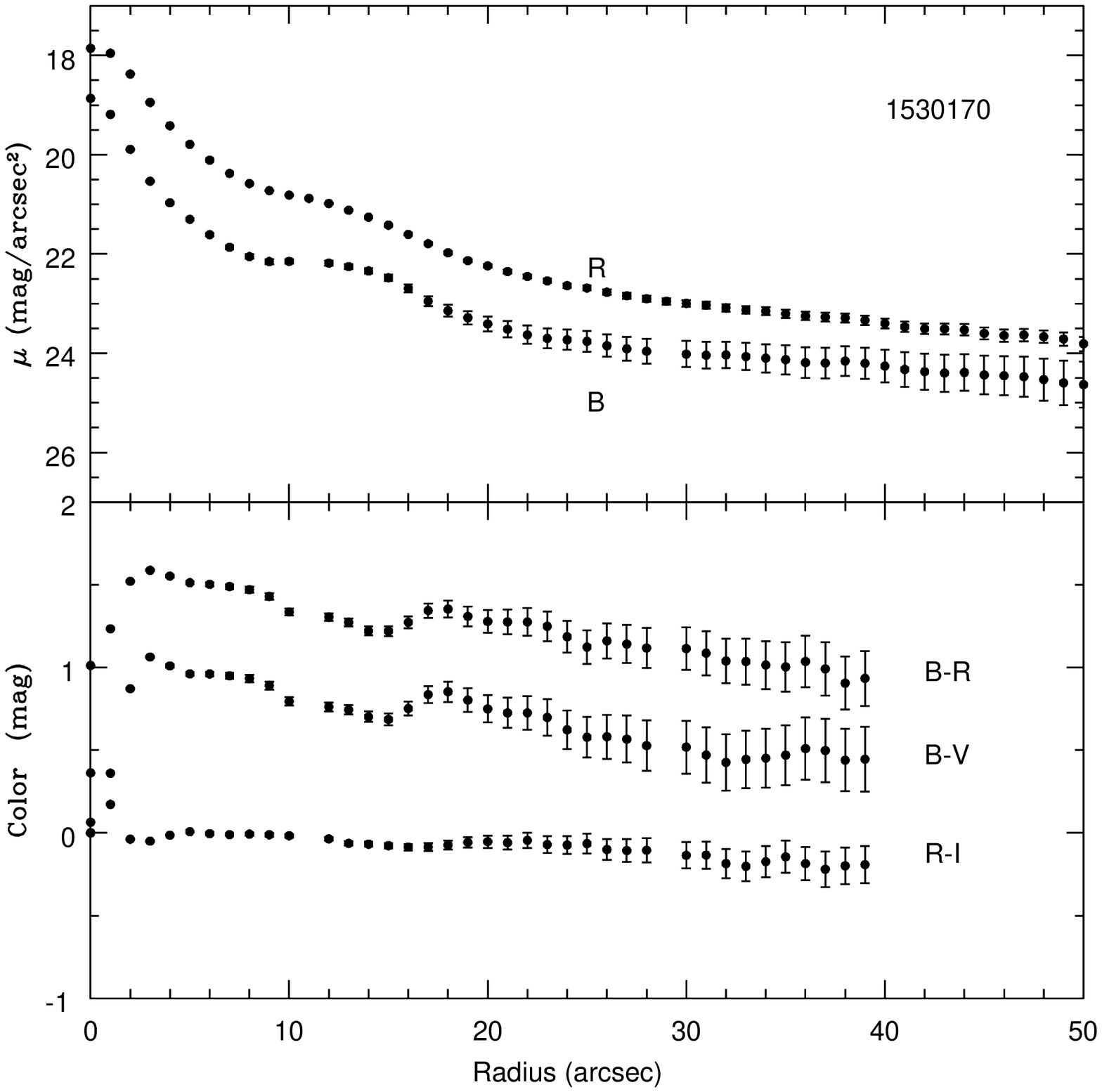}}
\caption[]{(d) Top panels contain \emph{B} and \emph{R} surface brightness profiles;
bottom panels contain radial color profiles for the galaxies
\object{ESO-LV 2060140}, \object{ESO-LV 5450360} and \object{ESO-LV
1530170}. In all cases \emph{R-I}
has been offset by -0.5 mag.}
\end{figure}
\begin{figure}[!t]
\setcounter{figure}{0}
\resizebox{\hsize}{!}{\includegraphics[height=5.6cm, width=8cm]{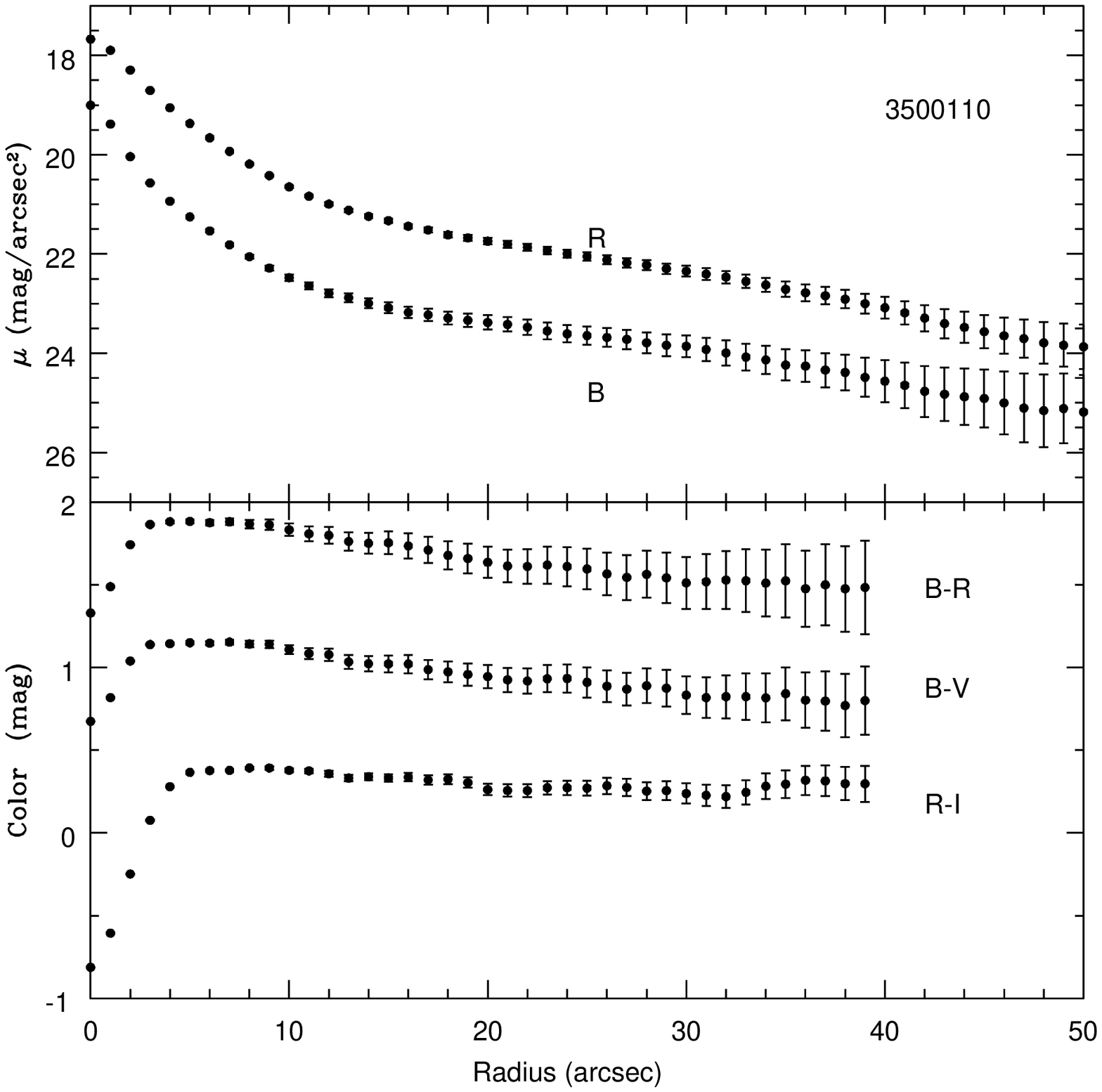}}
\resizebox{\hsize}{!}{\includegraphics[height=5.6cm, width=8cm]{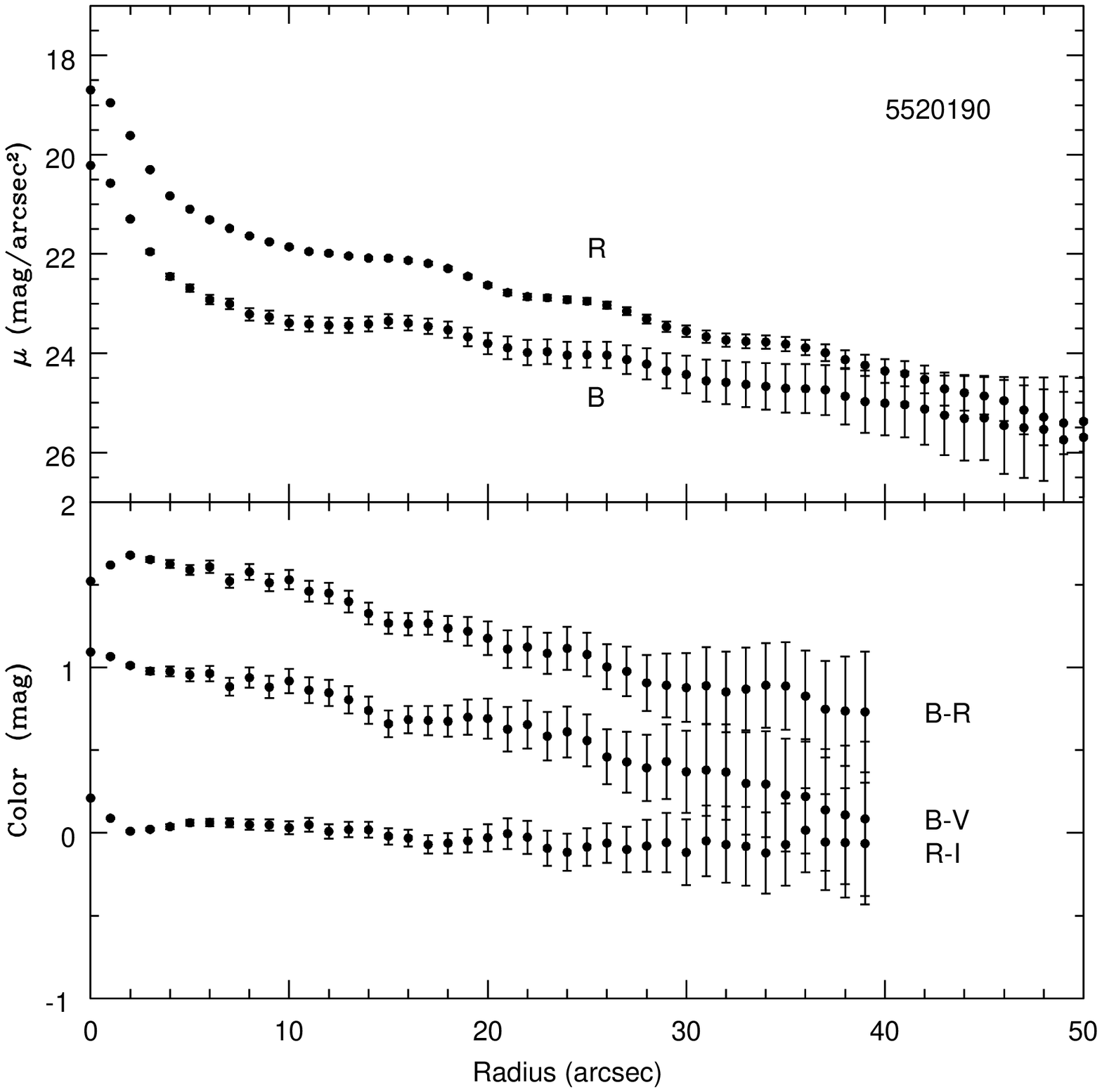}}
\resizebox{\hsize}{!}{\includegraphics[height=5.6cm, width=8cm]{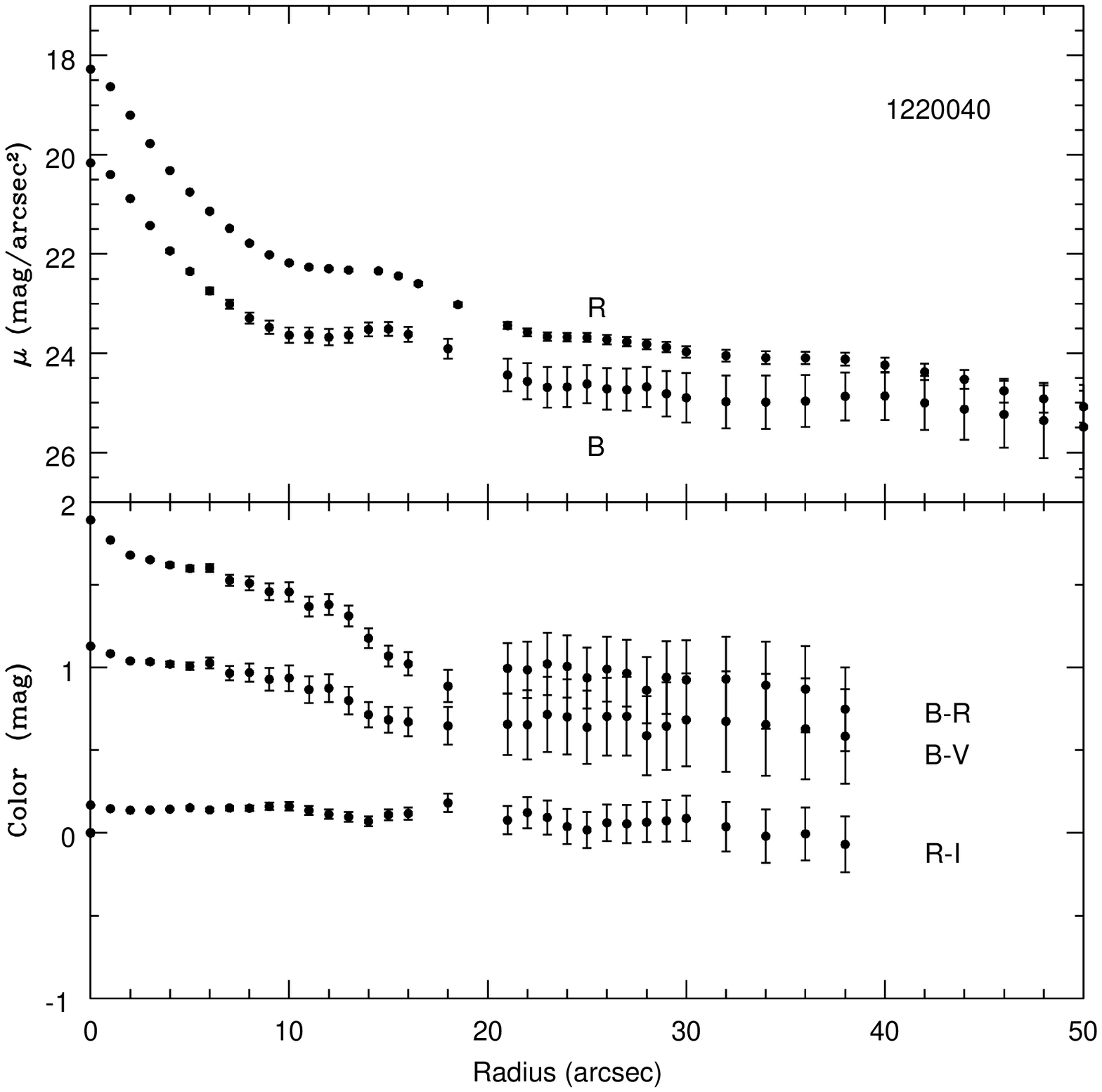}}
\caption[]{(e) Top panels contain \emph{B} and \emph{R} surface brightness profiles;
bottom panels contain radial color profiles for the galaxies
\object{ESO-LV 3500110}, \object{ESO-LV 5520190} and \object{ESO-LV
1220040}. In all cases \emph{R-I}
has been offset by -0.5 mag.}
\end{figure}
\begin{figure}[!t]
\setcounter{figure}{0}
\resizebox{\hsize}{!}{\includegraphics[height=5.6cm, width=8cm]{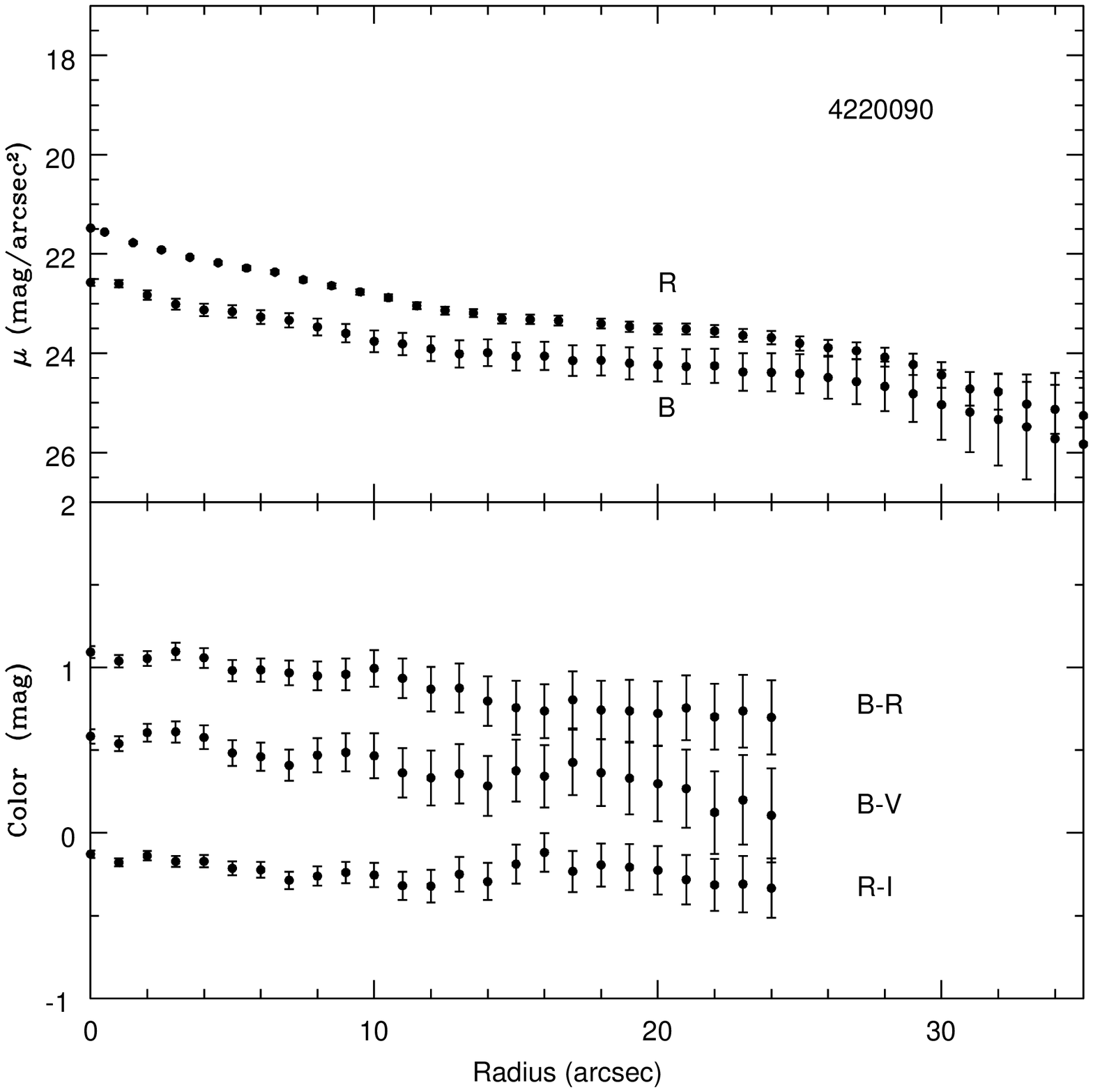}}
\resizebox{\hsize}{!}{\includegraphics[height=5.6cm, width=8cm]{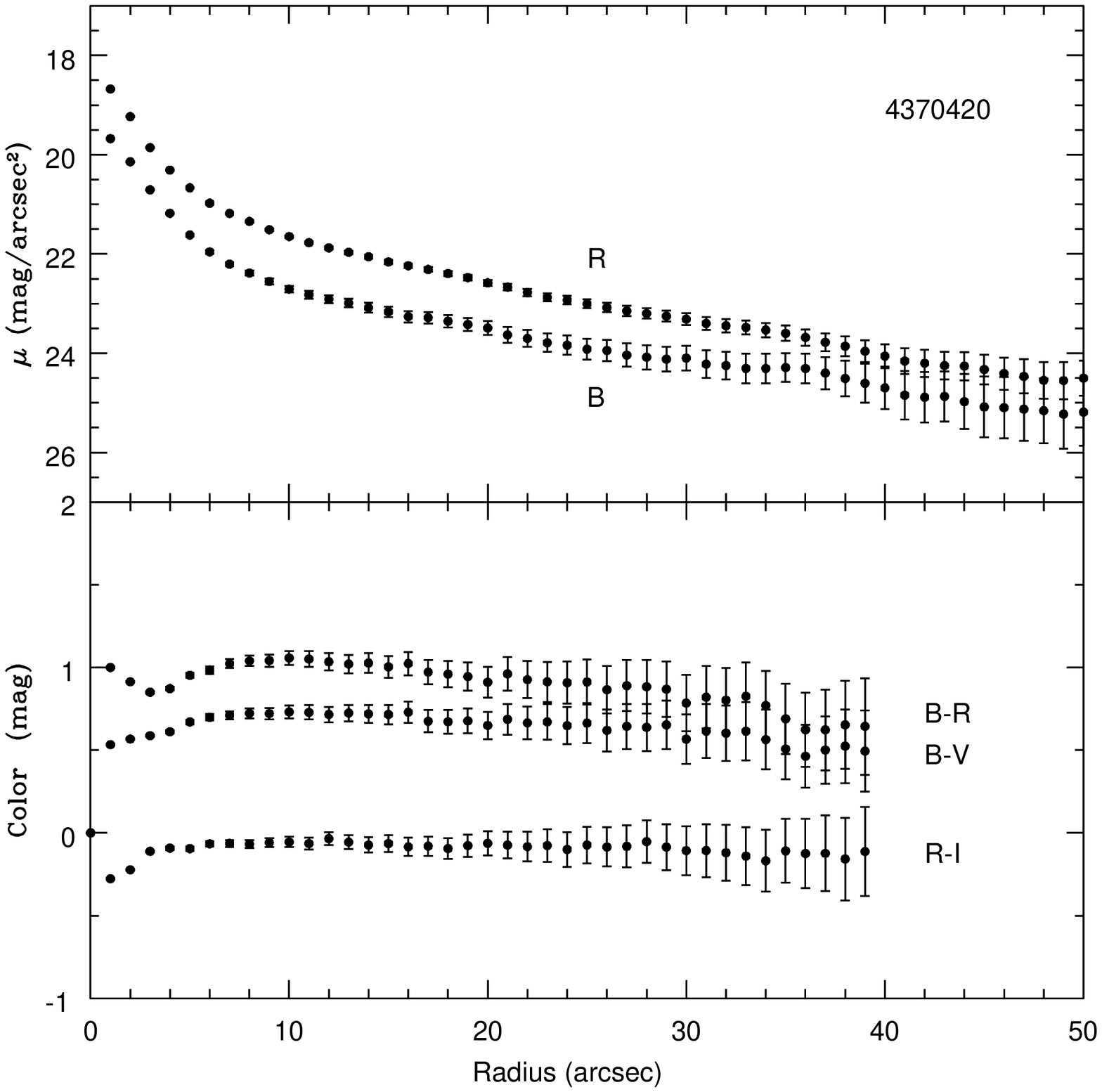}}
\resizebox{\hsize}{!}{\includegraphics[height=5.6cm, width=8cm]{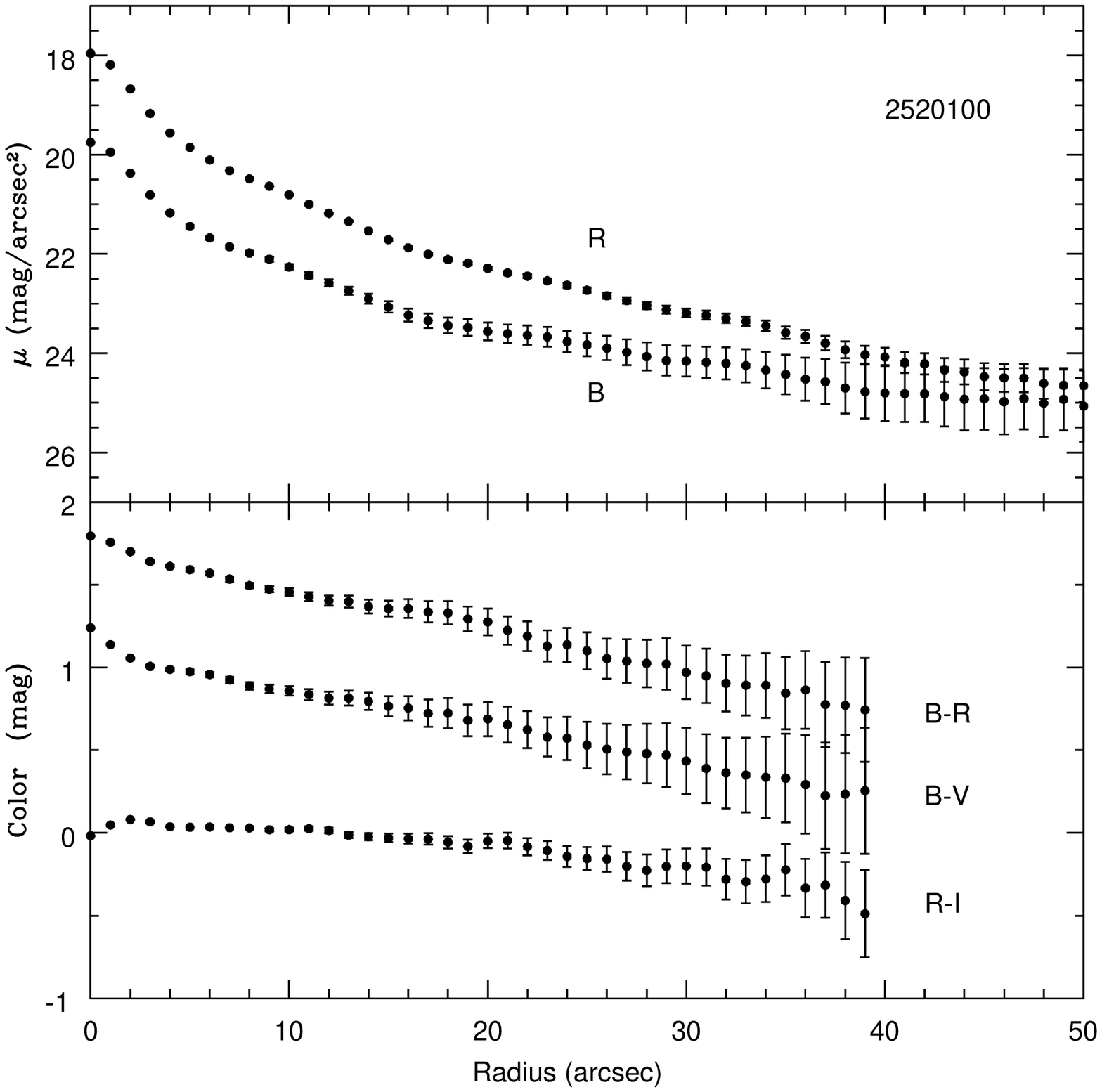}}
\caption[]{(f) Top panels contain \emph{B} and \emph{R} surface brightness profiles;
bottom panels contain radial color profiles for the galaxies
\object{ESO-LV 4220090}, \object{ESO-LV 4370420} and \object{ESO-LV
2520100}. In all cases \emph{R-I}
has been offset by -0.5 mag.}
\end{figure}
\begin{figure}[!t]
\setcounter{figure}{0}
\resizebox{\hsize}{!}{\includegraphics[height=5.6cm, width=8cm]{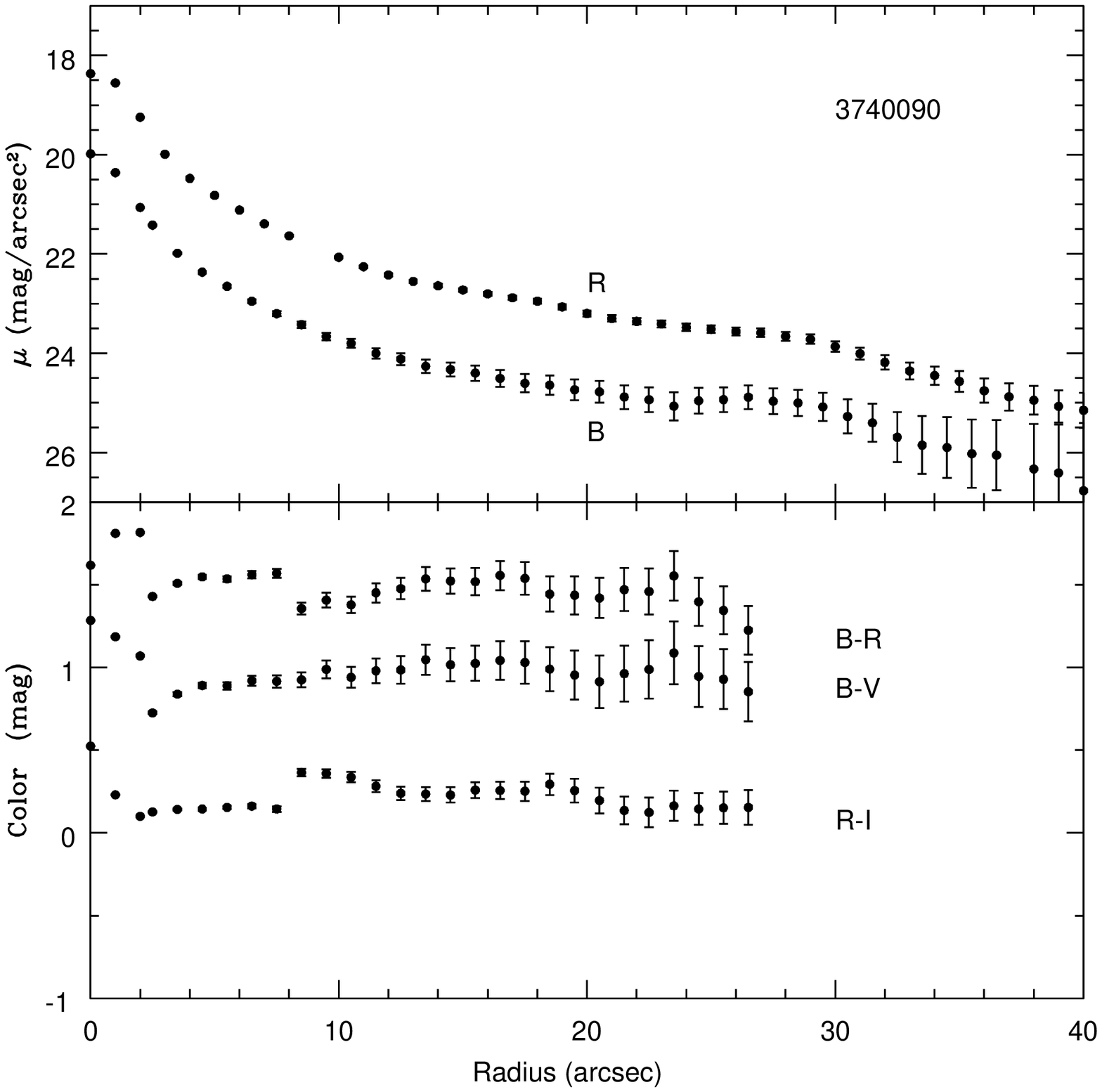}}
\caption[]{(g) Top panel contains \emph{B} and \emph{R} surface brightness profiles;
bottom panel contains radial color profiles for \object{ESO-LV
3740090}. \emph{R-I} has been offset by -0.5 mag.}  
\end{figure}
\begin{figure}
\resizebox{\hsize}{!}{\includegraphics{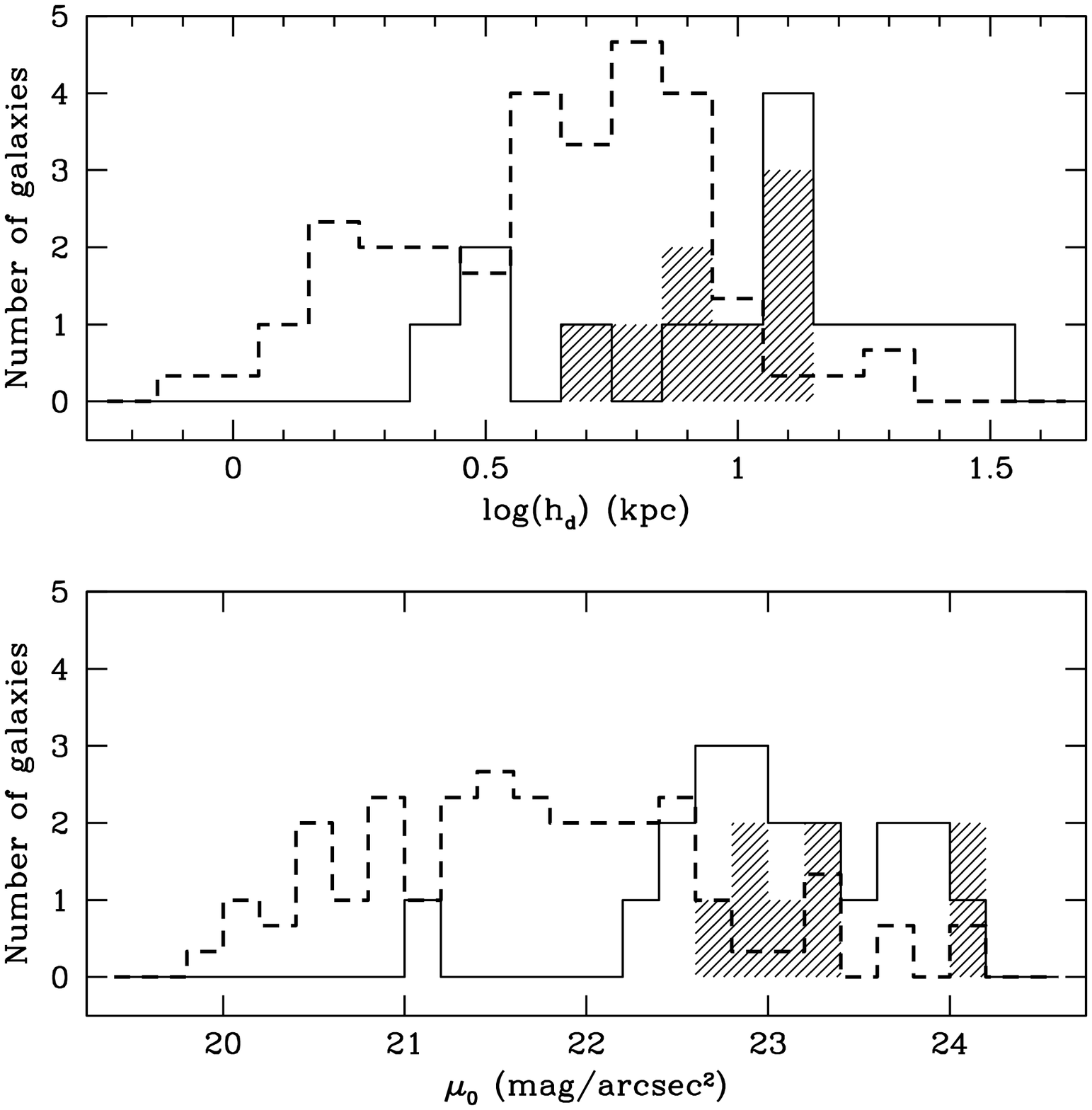}}
\caption[]{Histograms of the \emph{B} band structural disk parameters
$h_{\rm d}$
and $\mu_{0,\rm d}$. Solid line = bulge LSB galaxies. Dashed line = HSB
galaxies (dJ95). Shaded region = giant LSB galaxies (Sprayberry et
al. \cite{sprayberry}). The histogram of the dJ95 sample is divided by 3.
There is no preferred value of either $h_{\rm d}$ or $\mu_{0,\rm d}$. Note the HSB
galaxy in our sample.} 
\label{histogram}
\end{figure}

\section{Structural parameters}

\subsection{Central surface brightnesses and scale lengths}

\begin{figure}[!t]
\resizebox{\hsize}{!}{\includegraphics[height=5.6cm, width=8cm]{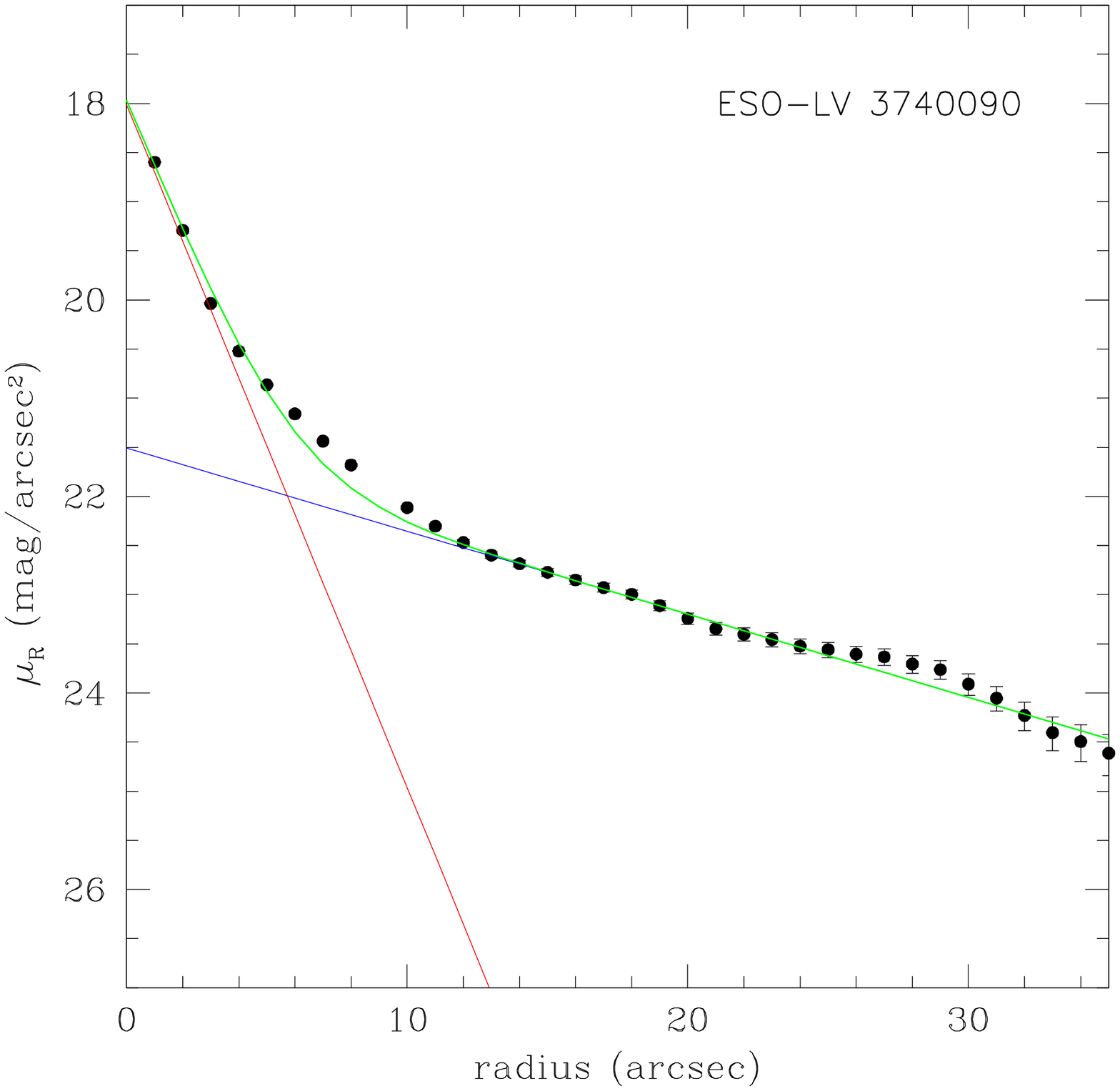}}
\resizebox{\hsize}{!}{\includegraphics[height=5.6cm, width=8cm]{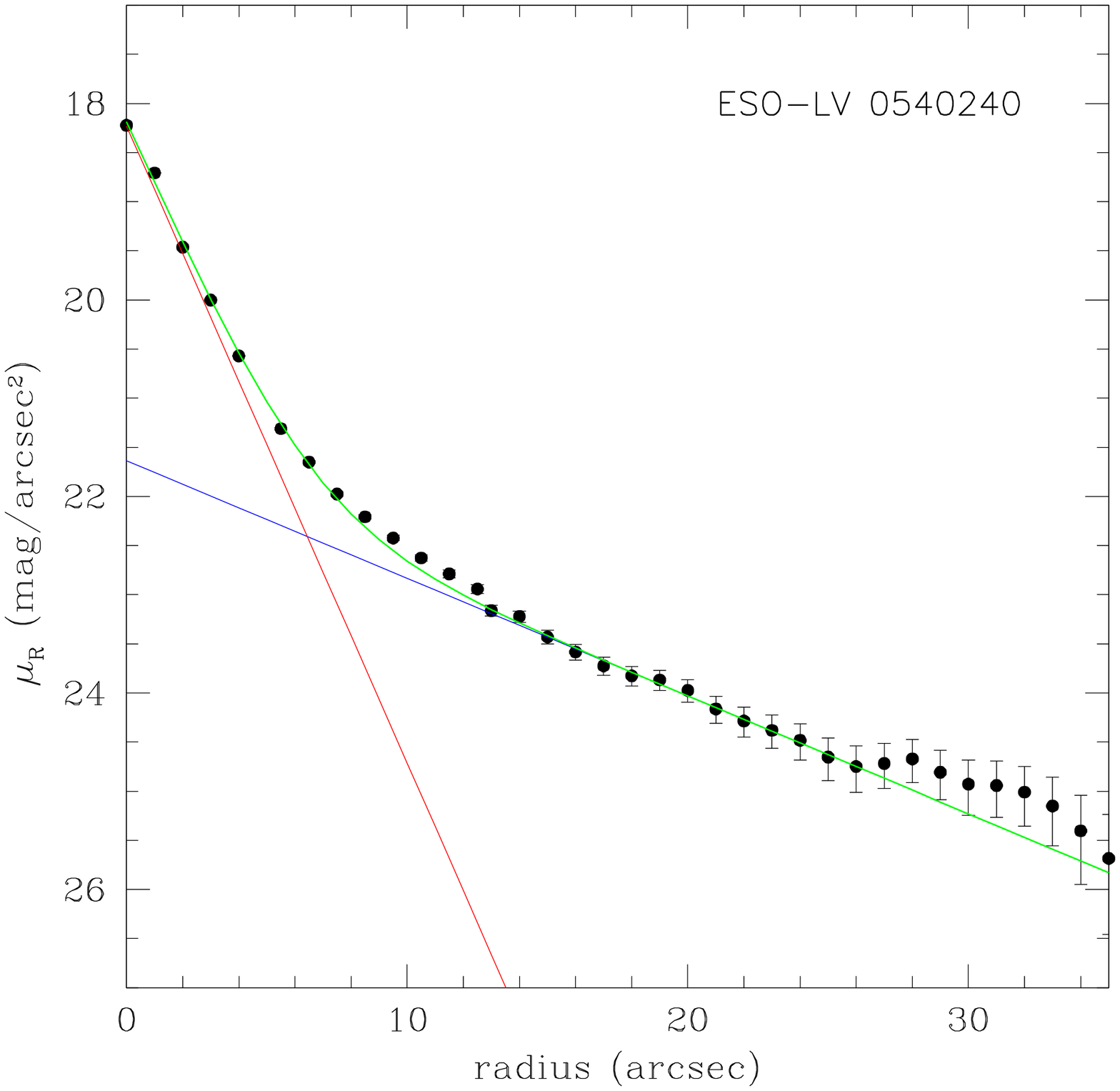}}
\caption[]{Bulge-disk decompositions. The exponential bulge and disk
fits plus the total profiles are shown. }
\label{fit}
\end{figure}

Most luminosity profiles clearly reveal the presence of a (HSB) bulge embedded in a low surface brightness disk
(Figs.~\ref{profiles}, upper panels). The profiles were decomposed into
bulge and disk components by taking \emph{exponential} profiles of the form

\begin{equation}
\Sigma(r)~=~\Sigma(0)~\mathrm{exp}(-r/h)^{1/n}
\label{formula_exponential}
\end{equation}

\noindent where $\Sigma(0)$ is the central intensity (e.g. L$_{\odot}$
pc$^{-2}$), $h$ is the scale length of either bulge or disk and $n$ is
taken to be 1 for both bulge and disk. Transforming to a logarithmic scale this equation becomes

\begin{equation}
\mu(r)~=~\mu_{0}~+~1.086~\times~{(r/h)}^{1/n}
\end{equation}

\noindent where $\mu_{0}$ is the central surface brightness \mbox{(mag
arcsec$^{-2}$)} of either disk or bulge and $r$ is the distance along the
major axis. Typical errors are 15\% in scale length and 0.1
mag arcsec$^{-2}$ in surface brightness.
In the past people used r$^{1/4}$ profiles for fitting bulges of
all types of spiral galaxies. But it is now known that the bulge
exponent correlates well with the morphological type of the
galaxy. Ellipticals and S0's have luminosity profiles that follow the
r$^{1/4}$ law (de Vaucouleurs \cite{de vaucouleurs}), while bulges of Sa and Sb galaxies are better described
by r$^{1/2}$ profiles. The bulges of later types can be well
represented by exponential ($n=1$) luminosity profiles (Andredakis
\cite{andredakis}; de Jong
\cite{de jong}, hereafter dJ95). However, the bulges of the galaxies in our sample
were \emph{all} better fitted with exponential profiles, regardless of
their
exact morphological types. This of course should
not come as a surprise given the large difficulties in assigning
Hubble types to such low contrast galaxies. Examples of bulge-disk decompositions are shown in
Figs.~\ref{fit}.

The structural
parameters derived from our fits are listed in
Table~\ref{structural_parameters}. The histograms of the structural
disk parameters, scale length and central surface brightness, for our
sample, a HSB sample (dJ95) and a giant LSB sample (Sprayberry et
al. \cite{sprayberry}) are
shown in Fig.~\ref{histogram}.

The distribution of \emph{B} band disk scale lengths is very broad
and has a median value of $h_{\rm d}$ = 12.6 kpc. Our LSB sample has a
median disk central surface brightness of
$\mu_{0,\rm d}$ = 23.04 \emph{B} mag arcsec$^{-2}$. There is no preferred value of
either. The one HSB
galaxy (\object{ESO-LV 1150280}) in our sample (Fig.~\ref{histogram}) was not used in all further discussions
and conclusions about bulge dominated LSB galaxies. The fact that
there is only one HSB disk galaxy in our sample confirms that our
selection was succesful in picking LSB disks.

\subsection{Effect of seeing on bulge scale lengths}

The typical sizes of
the bulges in our sample are $\sim 10 \arcsec$ and the mean \emph{B}
band bulge scale length is 2.37\arcsec. The bulges are thus 
larger than the seeing disk ($\sim 1.5
\arcsec$) and only the innermost ($< 2\arcsec$) parts are not
resolved. In order to investigate the effect of seeing on the fitted
bulge scale lengths we used the images of the galaxies with the
smallest bulges as scaled models and smoothed these to various
resolutions. It appeared that only the innermost one or two points are
affected. The fitted scale lengths, which are largely determined by
the outerpoints between radii of 2\arcsec and $\sim 10\arcsec$, do not
change within the stated errors, so we did not attempt to correct for seeing.

\subsection{Total magnitudes}

Total apparent magnitudes can be determined by integrating
Eq.~\ref{formula_exponential} out to infinity and converting to a
logarithmic scale: 
\begin{equation}
m_{\rm T}~=~\mu_{0}~-~2.5~log~(2\pi h^2)
\label{formula_magnitudes}
\end{equation}

\begin{figure}[!t]
\resizebox{\hsize}{!}{\includegraphics{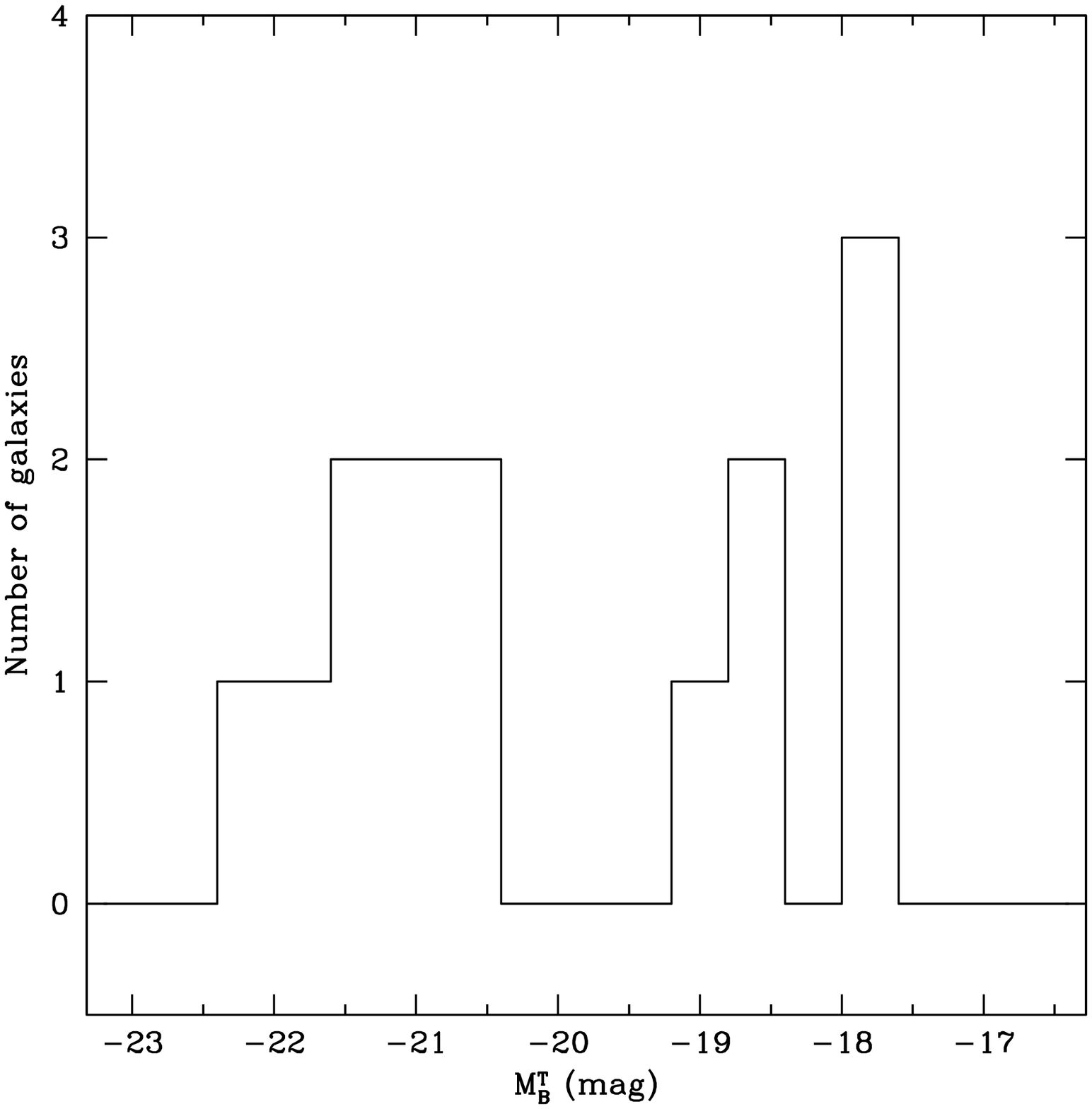}}
\caption[]{Histogram of the total absolute \emph{B} magnitudes. LSB galaxies
can be very luminous.} 
\label{maghisto}
\end{figure}

\noindent Optical depth effects in the galaxies have not been
taken into account, because of the small inclinations. However, we did
not use Eq.~\ref{formula_magnitudes} as $m_{\rm T}$ depends on
extrapolation of the light profile which introduces
uncertainties. Furthermore, it does not take into account bars and
rings which are clearly present in some of the galaxies in our sample. To
determine total absolute magnitudes we used aperture magnitudes
$m_{\rm ap}$ instead. We summed all the light to $2\sigma$ above sky
($\sim 25.5$ \emph{B} mag arcsec$^{-2}$) and
converted this
to magnitudes. We used the distances listed in
Table~\ref{sample_galaxies} to derive total absolute
magnitudes. Figure~\ref{maghisto} shows the distribution of absolute magnitudes
$M_{\rm{B}}^{\rm{T}}$ derived in this way. It is evident that low
surface brightness does not imply low luminosity. LSB galaxies can be
very luminous.

\subsection{Colors and color profiles}

Radial color profiles were obtained by subtracting two radial
luminosity profiles from each other. The bottom panels of Figs.~\ref{profiles}  show 
color profiles of the colors \emph{B-R}, \emph{B-V} and
\emph{R-I}. The errors in the color profiles were estimated by adding
in quadrature the errors in the surface brightness profiles. Most galaxies show color gradients and
become bluer with increasing radius (declining profiles). It is believed that the dust
content in disk LSB galaxies is low (dB95). Despite having no CO and
metallicity information of the galaxies in our sample we probably can assume
that some bulge LSB galaxies are also dust poor - in \object{ESO-LV 4250180} you can actually see a background
galaxy through the spiral arms. Then these color gradients are probably not
caused by dust only. The bluer colors in the outer parts could be
caused by relatively more light coming
from a younger population at larger radii. This suggests that star formation started in
the inner parts and progressed outwards.

Total integrated colors were determined by dividing sky-subtracted
images and then integrating within ellipses. The same ellipse parameters as
for determining the radial luminosity profiles have been used. We
determined three total colors. The nuclear color is the color of the
inner 10\arcsec. The luminosity weighted color is the color
as measured through a large isophotal aperture covering almost the
entire galaxy. The area weighted color is determined from the
average of the colors of many pixel-sized apertures over the
entire disk. When determining luminosity weighted colors most
weight is given to bright regions (nucleus, \ion{H}{ii} regions) of the galaxy. This is in contrast
with area weighted colors where all parts have equal weight and only
the area matters. Table~\ref{colors} gives total colors of our
sample. The estimated errors are $\sim 0.1$ mag.

\section{Discussion}

\subsection{Structural parameters}

Here and in the next subsection we will discuss the structural
parameters and colors of bulge dominated LSB galaxies and compare them
to disk dominated LSB galaxies and HSB galaxies. We will focus on trends to
explore the question whether bulge dominated LSB galaxies fit in with
the general trends defined by HSB galaxies, and more importantly,
whether they form the ``missing link'' between HSB and giant LSB
galaxies.
 
The majority of disk dominated LSB and HSB galaxies has disk scale lengths between 2 and 6 kpc (McGaugh
\& Bothun \cite{mc gaugh}; de Blok
\cite{de blok}; dJ95). The galaxies in our sample
have much larger disk scale lengths and the largest galaxies also have
bulges.  There appears a trend that the longest disk
scale lengths appear in galaxies with the longest bulge scale
lengths. 

\begin{figure}[!t]
\resizebox{\hsize}{!}{\includegraphics{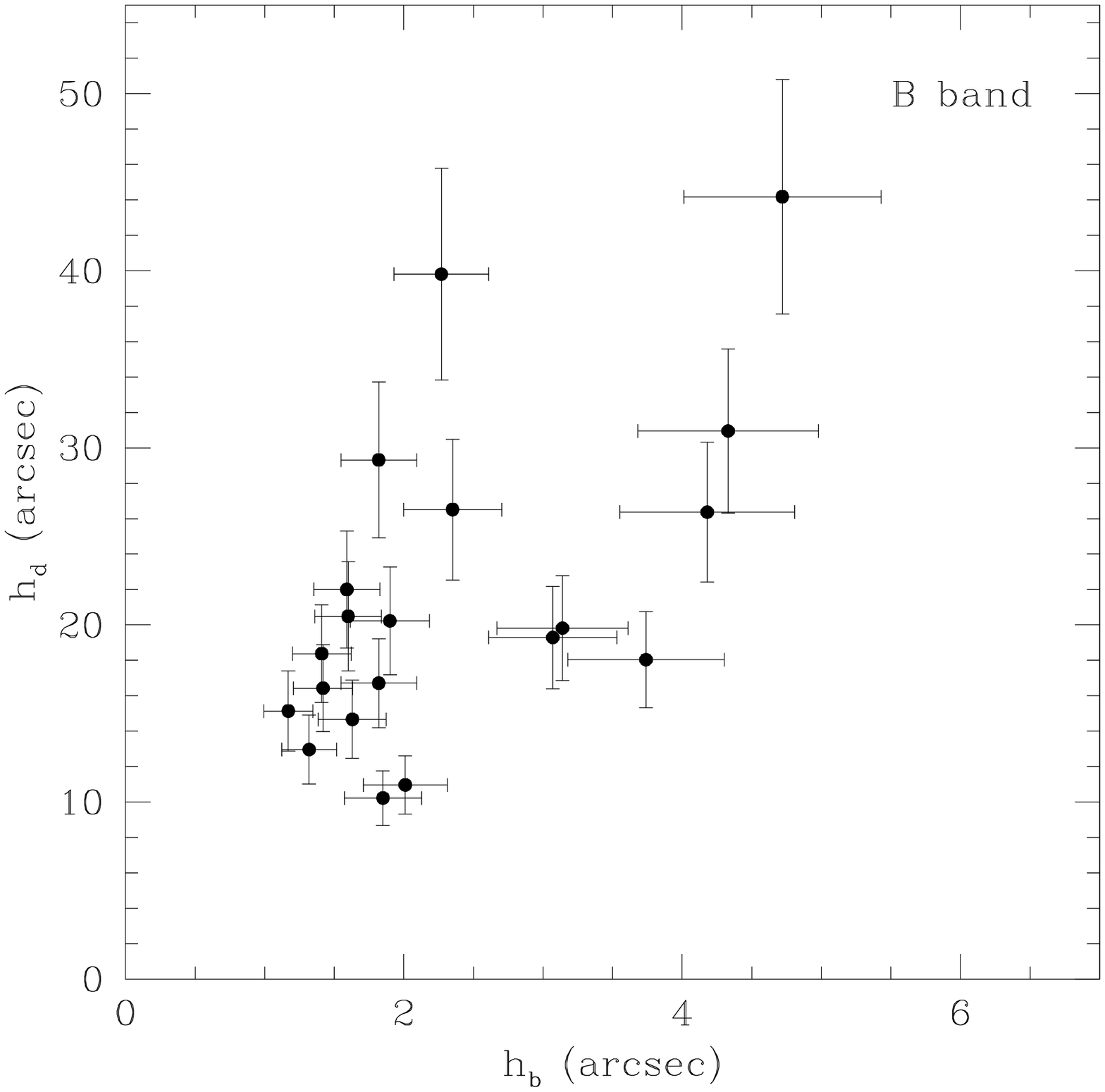}}
\caption[]{The \emph{B} band distribution of disk scale length with bulge
scale length. There is a correlation at 99.8\% probability which
suggests a coupling in the formation of bulge and disk.}
\label{bdcorrel}
\end{figure}

We use the Pearson correlation coefficient to determine the
significance of the correlation and find r = 0.59. We thus find a
restricted range for the bulge-to-disk ratios at the 
99.8\% probability level. The correlation between \emph{B} band disk and bulge scale
length is shown in Fig.~\ref{bdcorrel} and is consistent with other
studies (dJ95; Courteau et al. \cite{courteau}).

However, the trend could be partly due to the
selection criteria used, because we are discriminating against large
bulge, small disk galaxies.
 Galaxies with compact, bright bulges and
faint extended disks would comply to the criteria, but we do not have
them in our sample. Furthermore, there are no 
large, pure disk, LSB galaxies known till now, so it is likely that only the area below
the plotted trend is affected by selection effects. A correlation between disk and bulge scale length suggests that the formation of bulge and disk is
coupled.

\begin{figure}[!t]
\resizebox{\hsize}{!}{\includegraphics{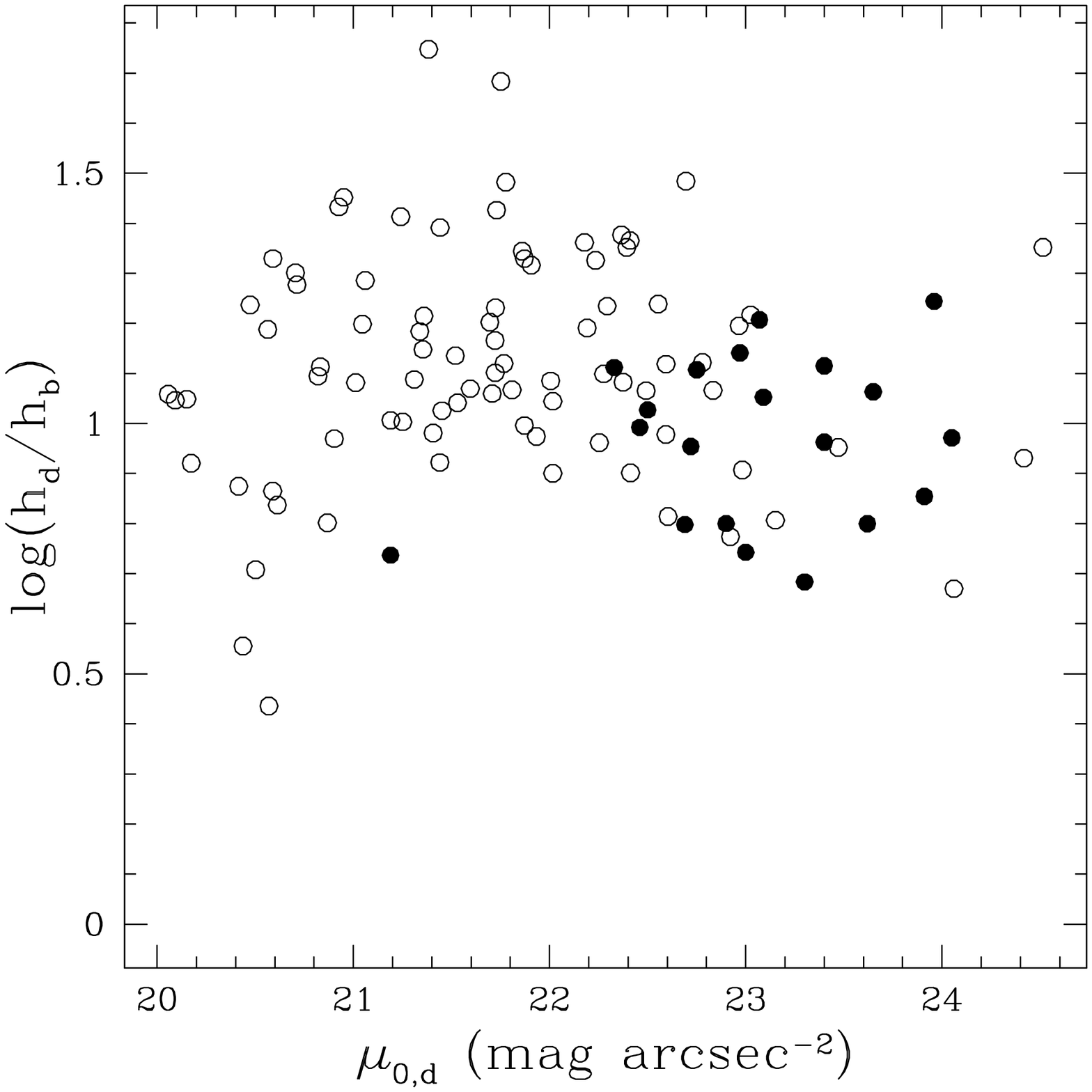}}
\caption[]{Distribution of disk-to-bulge scale length ratio of 
LSB and HSB galaxies with \emph{B} band
$\mu_{0,\rm d}$. Open circles = HSB galaxies (dJ95). Filled circles = LSB galaxies. LSB galaxies follow the trend defined by the HSB galaxies.}
\label{sclratios}
\end{figure}

The ratio of disk-to-bulge scale length for both HSB and LSB
galaxies has large scatter around $\sim 10$ and is illustrated in
Fig.~\ref{sclratios}. The LSB galaxies continue the trend defined by HSB
galaxies towards lower surface brightnesses.

\begin{table*}
\caption{Structural parameters of bulge dominated LSB galaxies.}
\begin{flushleft}
\begin{tabular}{rlllrlccrlllrlcc}
\hline
(1)&(2)&(3)&(4)&(5)&(6)&(7)&(8)&(1)&(2)&(3)&(4)&(5)&(6)&(7)&(8)\\
Name&Band&$\mu_{0,\rm d}$&$\mu_{0,\rm b}$&$h_{\rm d}$&$h_{\rm
b}$&$h_{\rm d}^\prime$&$h_{\rm b}^\prime$&Name&Band&$\mu_{0,\rm
d}$&$\mu_{0,\rm b}$&$h_{\rm d}$&$h_{\rm b}$&$h_{\rm d}^\prime$&$h_{\rm
b}^\prime$\\
\hline
0140040&B&22.46&20.18&12.96&1.32&-&-&2060140&B&22.97&21.75&22.00&1.59&21.3&1.5\\
&V&21.71&19.12&14.31&1.53&-&-&&V&22.14&20.93&18.90&1.76&18.3&1.7\\
&R&20.67&18.32&9.60&1.29&-&-&&R&21.57&20.19&17.75&1.62&17.2&1.6\\
&I&19.98&17.69&8.96&1.25&-&-&&I&20.92&19.83&15.97&1.77&15.5&1.7\\
1590200&B&22.90&22.28&19.82&3.14&-&-&1150280&B&21.19&19.71&10.97&2.01&4.6&0.8\\
&V&22.40&21.58&18.35&3.49&-&-&&V&20.07&18.46&9.31&1.70&3.9&0.7\\
&R&21.92&20.89&18.22&3.47&-&-&&R&20.07&17.36&12.01&1.70&5.0&0.7\\
&I&21.45&20.37&18.46&2.76&-&-&&I&19.25&16.94&10.69&1.80&4.5&0.8\\
3740090&B&23.40&19.98&16.71&1.82&3.0&0.3&1530170&B&23.07&18.96&29.32&1.82&12.5&0.8\\
&V&22.28&18.73&15.18&1.70&2.7&0.3&&V&22.20&18.78&22.24&3.24&9.5&1.4\\
&R&21.37&18.32&12.07&1.83&2.2&0.3&&R&21.63&17.79&23.74&2.59&10.1&1.1\\
&I&20.69&17.37&11.96&1.59&2.2&0.3&&I&20.97&16.93&20.14&2.14&8.6&0.9\\
4350310&B&22.33&19.55&15.14&1.17&-&-&2520100&B&22.69&20.06&19.29&3.07&12.5&2.0\\
&V&21.27&18.62&11.38&1.23&-&-&&V&21.40&18.76&13.33&2.64&8.6&1.7\\
&R&20.72&17.88&10.81&1.17&-&-&&R&20.68&18.06&12.94&2.63&8.4&1.7\\
&I&20.14&17.18&10.75&1.10&-&-&&I&19.72&17.55&10.04&2.51&6.5&1.6\\
0050050&B&23.31&22.80&18.04&3.74&-&-&3500110&B&22.72&19.26&14.67&1.63&14.2&1.6\\
&V&22.52&21.66&14.09&2.21&-&-&&V&21.37&18.46&11.99&2.14&11.6&2.1\\
&R&22.09&21.11&13.93&2.41&-&-&&R&20.54&17.60&11.94&2.07&11.6&2.0\\
&I&21.63&20.53&12.64&2.49&-&-&&I&20.23&17.83&14.14&4.47&13.7&4.3\\
0540240&B&23.00&19.68&10.23&1.85&9.7&1.8&4220090&B&24.05&23.13&44.18&4.72&13.4&1.4\\
&V&22.42&18.84&11.58&1.72&11.0&1.6&&V&23.02&22.71&17.31&5.24&4.7&1.6\\
&R&21.65&18.18&13.09&1.82&12.4&1.7&&R&22.84&21.90&30.32&4.18&9.2&1.3\\
&I&22.13&17.63&15.73&1.85&14.9&2.7&&I&22.50&21.45&22.25&4.94&6.8&1.5\\
5650160&B&23.41&19.95&18.38&1.41&-&-&4250180&B&23.65&20.46&16.43&1.42&7.2&0.6\\
&V&22.31&18.88&14.64&1.44&-&-&&V&22.83&19.89&20.74&1.53&9.1&0.7\\
&R&21.90&18.21&14.86&1.69&-&-&&R&22.30&19.30&17.08&1.57&7.5&0.7\\
&I&21.36&17.63&13.18&1.73&-&-&&I&21.66&19.00&15.94&1.63&7.0&0.7\\
0590090&B&23.09&22.67&26.52&2.35&2.6&0.2&4990110&B&23.91&22.59&30.95&4.33&5.6&0.8\\
&V&22.19&21.46&25.62&2.48&2.5&0.2&&V&23.41&22.02&28.77&5.07&5.2&0.9\\
&R&21.57&20.78&24.99&2.89&2.4&0.3&&R&22.91&21.59&26.80&5.10&4.9&0.9\\
&I&20.87&20.29&21.13&3.28&2.0&0.3&&I&22.37&21.10&26.09&4.93&4.7&0.9\\
4370420&B&22.50&19.17&20.23&1.90&3.4&0.3&5450360&B&23.62&22.74&26.37&4.18&29.5&4.7\\
&V&22.00&18.73&21.10&2.16&3.5&0.4&&V&21.66&18.98&11.84&1.63&13.2&1.8\\
&R&21.58&18.29&18.26&1.95&3.1&0.3&&R&20.80&17.94&9.72&1.36&10.9&1.5\\
&I&21.02&18.04&17.15&2.02&2.9&0.3&&I&20.19&17.37&10.30&1.47&11.5&1.6\\
4400490&R&21.64&19.80&17.99&3.06&2.6&0.4&5520190&B&22.75&20.25&20.49&1.60&16.2&1.3\\
1220040&B&23.96&20.09&39.81&2.27&24.5&1.4&&V&21.49&19.16&12.90&1.50&10.2&1.2\\
&V&23.18&18.97&33.06&2.20&20.3&1.3&&R&20.86&18.78&12.32&1.75&9.7&1.4\\
&R&22.67&18.28&25.20&2.07&15.5&1.3&&I&20.33&18.04&11.98&1.54&9.5&1.2\\
&I&21.89&17.59&21.83&2.04&13.4&1.3&&&&&&&&\\
\hline
\multicolumn{8}{l}{\emph{Notes:}}\\
\multicolumn{8}{l}{(1) ESO-LV name of the galaxy.}\\
\multicolumn{8}{l}{(2) Photometric band.}\\
\multicolumn{8}{l}{(3) Central disk surface brightness in mag arcsec$^{-2}$.}\\
\multicolumn{8}{l}{(4) Central bulge surface brightness in mag
arcsec$^{-2}$.}\\
\multicolumn{8}{l}{(5) Exponential disk scale length in arcseconds.}\\
\multicolumn{8}{l}{(6) Exponential bulge scale length in arcseconds.}\\
\multicolumn{8}{l}{(7) Exponential disk scale length in kpc.}\\
\multicolumn{8}{l}{(8) Exponential bulge scale length in kpc.}\\
\label{structural_parameters}
\end{tabular}
\end{flushleft}
\end{table*}

\begin{table*}
\caption[]{Total colors of bulge dominated LSB galaxies.}
\begin{flushleft}
\begin{tabular}{rllll|llll|llll}
&\multicolumn{4}{c}{nuc}&\multicolumn{4}{c}{lum}&\multicolumn{4}{c}{area}\\
\hline
Name&B-V&B-R&V-R&R-I&B-V&B-R&V-R&R-I&B-V&B-R&V-R&R-I\\
\hline
\hline
ESO-LV 2060140&0.67&1.13&0.46&0.49&0.50&0.90&0.40&0.43&0.44&0.83&0.36&0.45\\
1150280&0.92&1.56&0.62&0.50&0.67&1.45&0.66&0.33&0.60&0.99&0.58&0.46\\
1530170&0.87&1.45&0.58&0.51&0.49&1.16&0.67&0.34&0.50&1.01&0.52&0.35\\
2520100&0.98&1.60&0.62&0.54&0.58&1.13&0.55&0.38&0.62&1.08&0.57&0.41\\
350110&1.09&1.81&0.72&0.75&0.95&1.68&0.73&0.75&0.89&1.63&0.69&0.88\\
4220090&0.49&0.93&0.44&0.34&0.26&0.76&0.49&0.24&0.30&0.73&-&0.18\\
4250180&0.67&1.13&0.46&0.42&1.32&1.40&0.08&0.43&0.96&1.22&0.18&0.50\\
4990110&0.59&0.98&0.39&0.39&0.92&1.11&0.19&0.42&0.62&1.01&0.34&0.47\\
5450360&1.06&1.72&0.66&0.62&1.05&1.62&0.57&0.65&0.92&1.41&0.54&0.69\\
5520190&0.95&1.56&0.61&0.54&0.39&0.60&0.21&0.55&0.57&1.20&0.52&0.57\\
1220040&0.93&1.43&0.50&0.58&0.87&1.16&0.29&0.50&0.66&0.74&0.36&0.54\\
140040&0.98&1.53&0.55&0.54&0.96&1.28&0.32&0.42&0.74&1.19&0.48&0.45\\
1590200&0.55&1.04&0.49&0.39&0.32&0.89&0.57&0.41&0.43&0.72&0.43&0.48\\
3740090&0.91&1.44&0.52&0.56&1.06&1.37&0.31&0.54&0.88&1.43&0.40&-\\
4350310&0.84&1.44&0.60&0.54&0.55&1.06&0.51&0.48&0.49&0.70&0.46&0.50\\
50050&0.53&0.91&0.38&0.31&0.49&0.94&0.45&0.10&0.37&0.67&0.27&0.26\\
540240&0.84&1.42&0.58&0.57&0.94&1.44&0.50&0.59&0.70&1.08&0.42&0.64\\
5650160&0.96&1.56&0.60&0.58&0.84&1.36&0.52&0.49&0.72&1.17&0.50&0.51\\
590090&0.69&1.12&0.43&0.40&0.61&0.99&0.38&0.30&0.59&0.96&0.37&0.43\\
4370420&0.58&0.83&0.25&0.34&0.63&0.84&0.21&0.35&0.53&0.74&0.14&0.45\\
\end{tabular}
\label{colors}
\end{flushleft}
\end{table*}

\begin{figure}[!t]
\resizebox{\hsize}{!}{\includegraphics{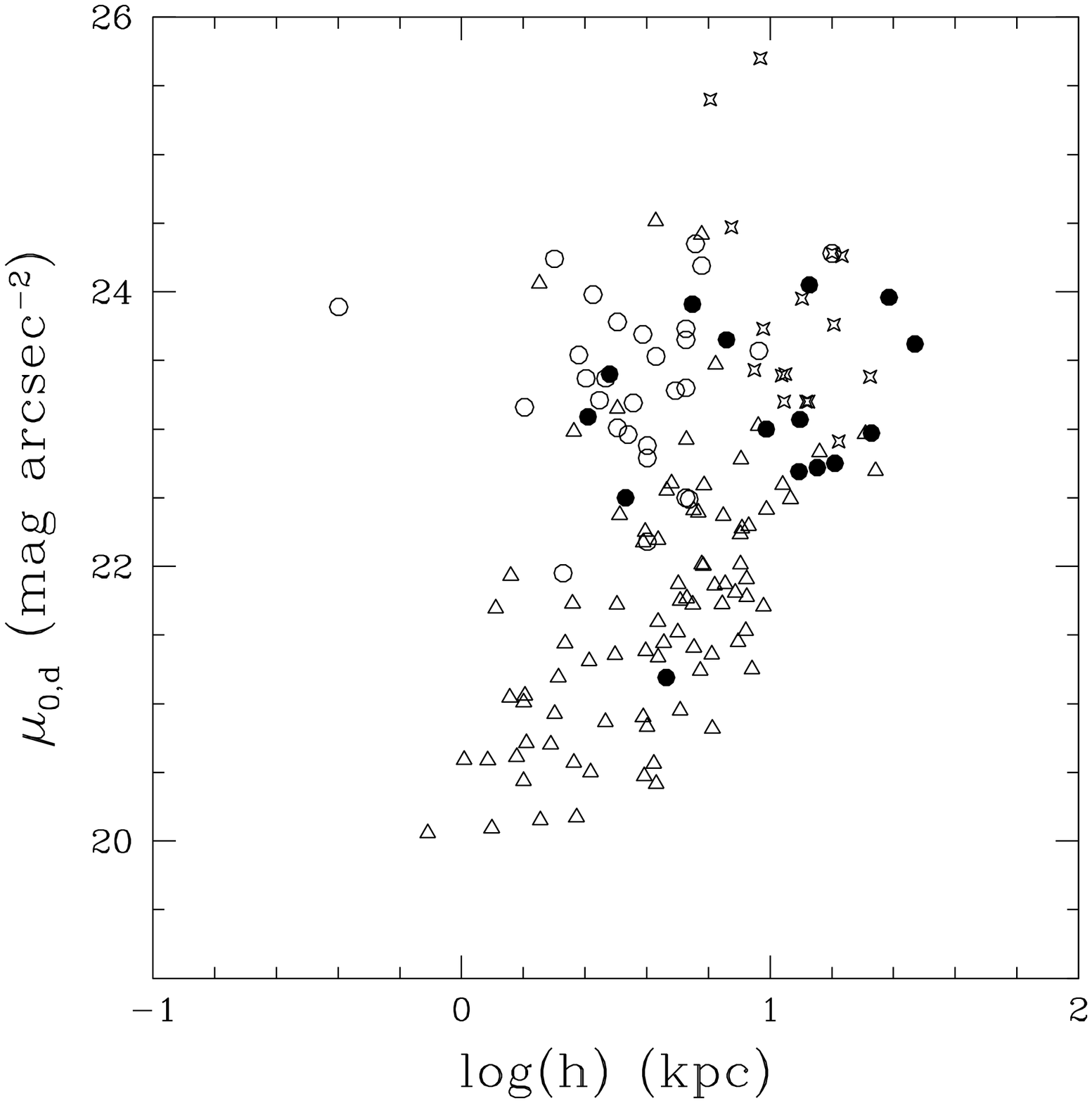}}
\caption[]{\emph{B} band disk central surface brightness versus disk scale
length. Triangles = HSB galaxies (dJ95). Open circles = disk LSB
galaxies (dB95). Stars = giant LSB galaxies (Sprayberry et al. \cite{sprayberry}). Filled circles =
bulge LSB galaxies.}
\label{hmu}
\end{figure}

In Fig.~\ref{hmu} we plot the distribution of disk central surface brightness with disk
scale length for the samples of dJ95, dB95, Sprayberry
and the current sample. All samples fit in with the general trend that there are no
galaxies with high central surface brightnesses and large disk scale
lengths. Some of the galaxies in our sample have such large disk scale
lengths that they are found in the region of giant LSB galaxies and therefore could be classified as such.

\begin{figure}[!t]
\resizebox{\hsize}{!}{\includegraphics{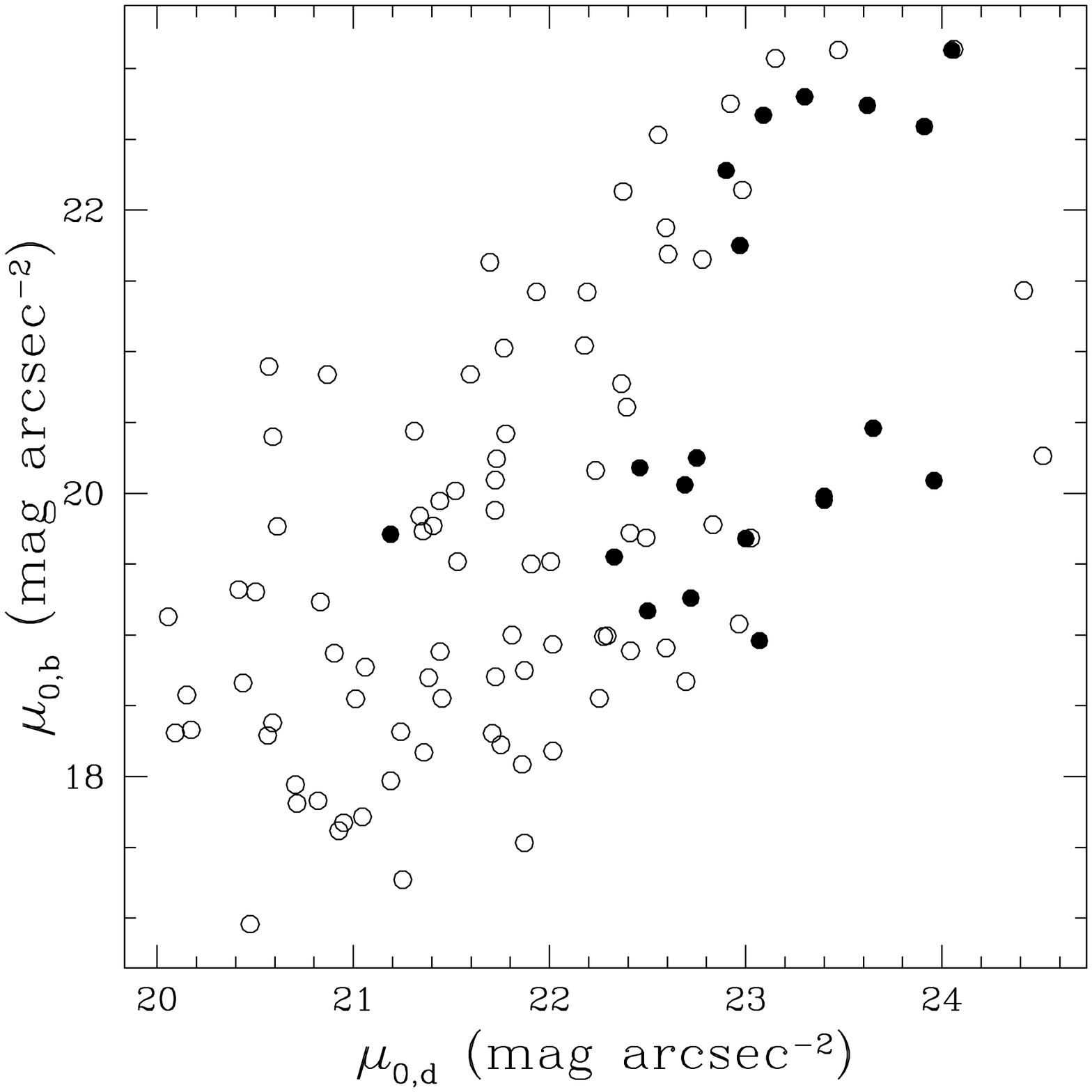}}
\caption[]{\emph{B} band central bulge surface brightnesses versus central disk
surface brightnesses. Open circles = HSB galaxies (dJ95). Filled
circles =  LSB galaxies. Note how the LSB galaxies split into two groups.}
\label{mumu}
\end{figure}

To investigate whether disk and bulge central surface brightnesses are
related we plot these parameters in Fig.~\ref{mumu} for the samples of
dJ95 and our sample. The figure clearly shows how the LSB galaxies
fill the low surface brightness region, but split into two groups; one
near $\mu_{0,\rm b}$ = 20
mag arcsec$^{-2}$
and one near $\mu_{0,\rm b}$ = 22 mag arcsec$^{-2}$. These groups do not
correspond to the two types of bulge LSB galaxies mentioned in Section
2, but the groups are a mixture of both these types.
There seems to be a very broad general tendency for both HSB and LSB galaxies to have brighter bulge central surface
brightnesses with increasing disk central surface brightnesses. 

\begin{table}
\caption[]{Bulge-to-disk ratios in the different bands.} 
\begin{flushleft}
\begin{tabular}{|l|l|r|c|l|l|r|}
\hline
ESO-LV \#&band&B/D&&ESO-LV \#&band&B/D\\
\hline
2060140&B&0.02&&140040&B&0.08\\
&V&0.03&&&V&0.12\\
&R&0.03&&&R&0.16\\
&I&0.03&&&I&0.16\\
1150280&B&0.13&&1590200&B&0.04\\
&V&0.15&&&V&0.08\\
&R&0.24&&&R&0.09\\
&I&0.24&&&I&0.06\\
1530170&B&0.17&&0050050&B&0.07\\
&V&0.50&&&V&0.05\\
&R&0.41&&&R&0.07\\
&I&0.47&&&I&0.11\\
2520100&B&0.29&&540240&B&0.59\\
&V&0.44&&&V&0.60\\
&R&0.46&&&R&0.93\\
&I&0.46&&&I&0.87\\
350110&B&0.30&&0590090&B&0.01\\
&V&0.47&&&V&0.02\\
&R&0.45&&&R&0.03\\
&I&0.91&&&I&0.04\\
4220090&B&0.03&&5650160&B&0.14\\
&V&0.12&&&V&0.23\\
&R&0.05&&&R&0.38\\
&I&0.13&&&I&0.53\\
4250180&B&0.14&&4370420&B&0.19\\
&V&0.08&&&V&0.21\\
&R&0.13&&&R&0.24\\
&I&0.12&&&I&0.21\\
4990110&B&0.07&&4350310&B&0.08\\
&V&0.11&&&V&0.13\\
&R&0.12&&&R&0.16\\
&I&0.11&&&I&0.16\\
5450360&B&0.06&&3740090&B&0.28\\
&V&0.22&&&V&0.33\\
&R&0.28&&&R&0.38\\
&I&0.27&&&I&0.38\\
5520190&B&0.06&&1220040&B&0.11\\
&V&0.12&&&V&0.21\\
&R&0.14&&&R&0.39\\
&I&0.14&&&I&0.46\\
\hline
\end{tabular}
\label{bdratios}
\end{flushleft}
\end{table}

\subsection{Colors}

The color profiles (Figs.~\ref{profiles}, bottom panels) show that in most galaxies the outer parts are bluer than the inner
parts. It was shown by dB95 that this is also true for late-type LSB
galaxies. It is hard to draw conclusions about the colors of
the bulges judging from the color profiles alone. But one can reveal
differences between disk and bulge colors by comparing
bulge-to-disk (B/D) ratios in different wavelength bands.
We determined B/D ratios by comparing the total light
output of bulge and disk per passband. We noticed a clear tendency for
the B/D ratios to increase towards the redder
wavelengths (Table~\ref{bdratios}). This means that the bulges of
LSB galaxies are \emph{redder} than their disks, confirming a trend
also observed for HSB galaxies (dJ95). 

In Table~\ref{total_colors} we present the integrated colors of the
galaxies in our sample
and those of disk dominated LSB (de Blok \cite{de blok}) and giant LSB galaxies
(Sprayberry et al. \cite{sprayberry}). Prior to determining the mean nuclear color
of the galaxies in our sample galaxies without a
clear bulge were excluded. For the determination of the mean area weighted
colors the galaxies without a clear bulge were included. The systems
in our sample have redder area weighted colors than disk LSB galaxies and this would be more pronounced if the more
or less bulgeless galaxies were left out of the sample. 

\begin{figure}[!t]
\resizebox{\hsize}{!}{\includegraphics{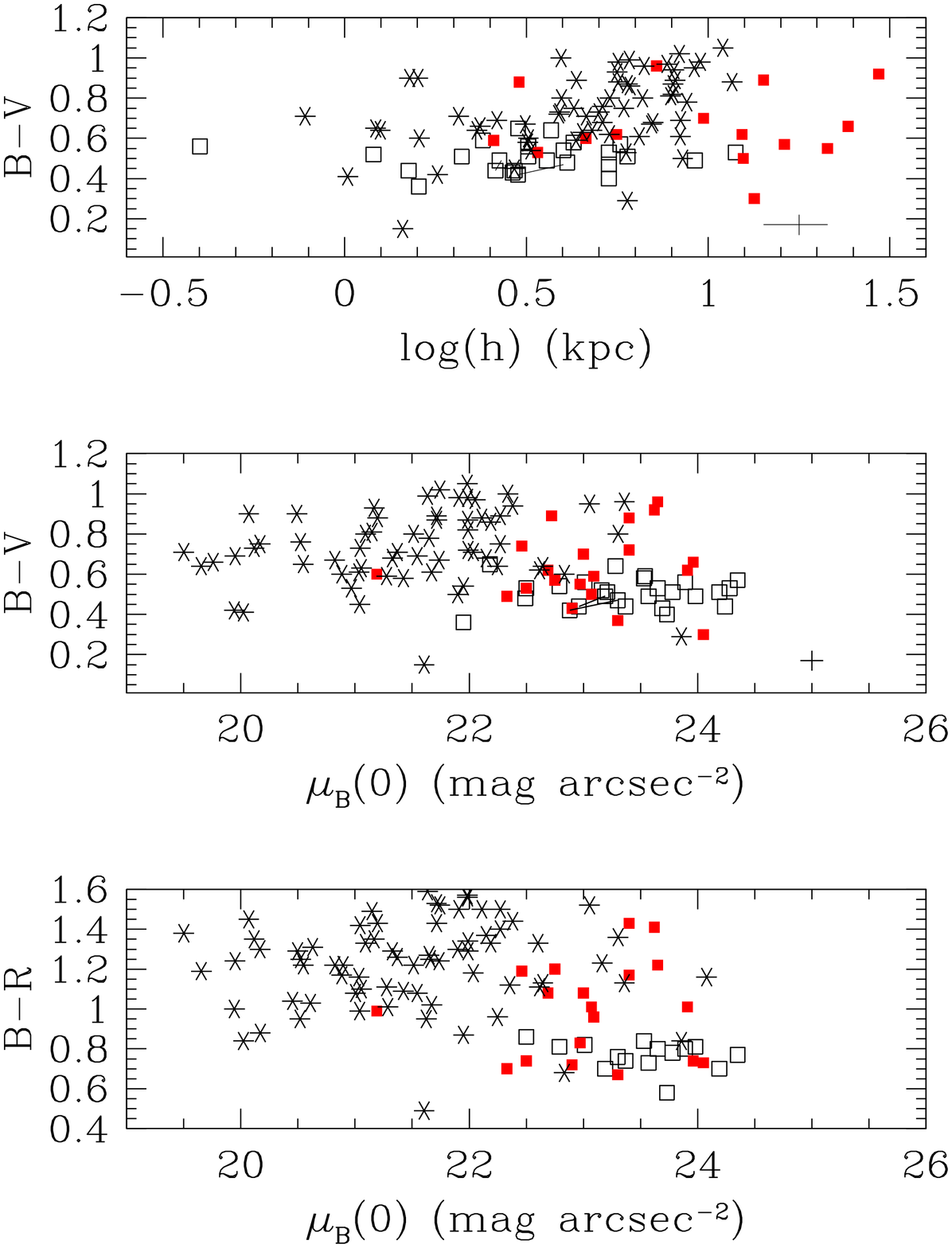}}
\caption[]{Top panel: \emph{B-V} versus disk scale length. Center
panel: \emph{B-V} versus \emph{B} $\mu_{0,\rm d}$. Bottom panel: \emph{B-R}
versus \emph{B} $\mu_{0,\rm d}$. Stars = HSB galaxies (dJ95). Open squares = disk LSB
galaxies (dB95). Filled squares = bulge LSB galaxies. Note how the bulge LSB
galaxies cover the entire range in (area weighted) color whereas the
disk LSB galaxies are concentrated in a rather blue area. } 
\label{mubv}
\end{figure}

Figure~\ref{mubv} shows the distribution of \emph{B-V} color with disk
scale length (top) as well as the distribution of \emph{B-V} color
(center) and \emph{B-R} color (bottom)
with disk central surface brightness. There is no trend of central
surface brightness with color and the colors do not depend on
size. HSB galaxies cannot be the progenitors of LSB galaxies since galaxies
fade and \emph{redden}. The bluest galaxies are the disk dominated LSB galaxies and are concentrated in a rather
small area whereas the HSB galaxies scatter over
the entire color range towards the redder colors. The bulge dominated
LSB galaxies fill up the region between these two samples. The
large scatter in color for the bulge dominated LSB sample is due to the wider range in
morphological types. The
disk dominated LSB galaxies are quiescent and form a fairly
uniform sample. The galaxies in our sample have red bulges, but some
of them also have bars and (blue) rings making the spread in color larger
than for a more uniform sample.     

Some of the large bulge dominated LSB galaxies in our sample could
easily be classified as giant LSB galaxies, judging from their sizes and luminosities. It is therefore instructive to compare the
colors with those of giant LSB galaxies. The disk colors of the biggest
galaxies in our sample are significantly \emph{bluer} than giant
disks (Table~\ref{total_colors}) and the B/D ratios
are smaller than for typical giants (Sprayberry et
al. \cite{sprayberry}). So although they have bulges and very large low surface
brightness disks, they are not as evolved as the giants.  

The comparison of our sample with a large sample of typical
``Freeman galaxies'' (de Jong \& van der Kruit \cite{de jong_vdkruit})
not only shows that on average the bulge dominated LSB galaxies are
bluer, but also that one can distinguish a LSB from a HSB galaxy of
the same type
by its bluer color.

\begin{table*}
\caption{Comparison of integrated colors.}
\begin{flushleft}
\begin{tabular}{lllllll}
\hline
\multicolumn{4}{c}{B-V}&\multicolumn{3}{c}{B-R}\\
\hline
(1)&(2)&(3)&(4)&(2)&(3)&(4)\\
Type&Nuc&Area&Disk&Nuc&Area&Disk\\
\hline
Bulge LSB&$0.87\pm0.15$&$0.63\pm0.19$&$0.54\pm0.30$&$1.31\pm0.29$&$1.03\pm0.27$&$0.88\pm0.40$\\
Disk LSB&$0.63\pm0.07$&$0.52\pm0.05$&$0.57\pm0.04$&$1.00\pm0.11$&$0.77\pm0.07$&$0.89\pm0.05$\\
Giant LSB&&&$0.73\pm0.05$&&&$1.23\pm0.06$\\
Largest galaxies&&&$0.4\pm0.2$&&&$0.8\pm0.4$\\
in our sample&&&&&&\\
\hline       
\multicolumn{3}{l}{\emph{Notes:}}\\
\multicolumn{3}{l}{(1) Type of galaxy.}\\
\multicolumn{3}{l}{(2) Mean nuclear colors.}\\
\multicolumn{3}{l}{(3) Mean area weighted colors.}\\
\multicolumn{3}{l}{(4) Mean disk colors.}\\
\end{tabular}
\label{total_colors}
\end{flushleft}
\end{table*}

\subsection{Bulge formation scenarios for LSB galaxies}

We will take a more detailed look at the two types of bulge LSB
galaxies as described in section 2 to see if any
systematic differences show up in their properties. We will also try
to make some rough quantitative statements about the bulges of the
galaxies in our sample. 

\begin{figure}[!t]
\resizebox{\hsize}{!}{\includegraphics{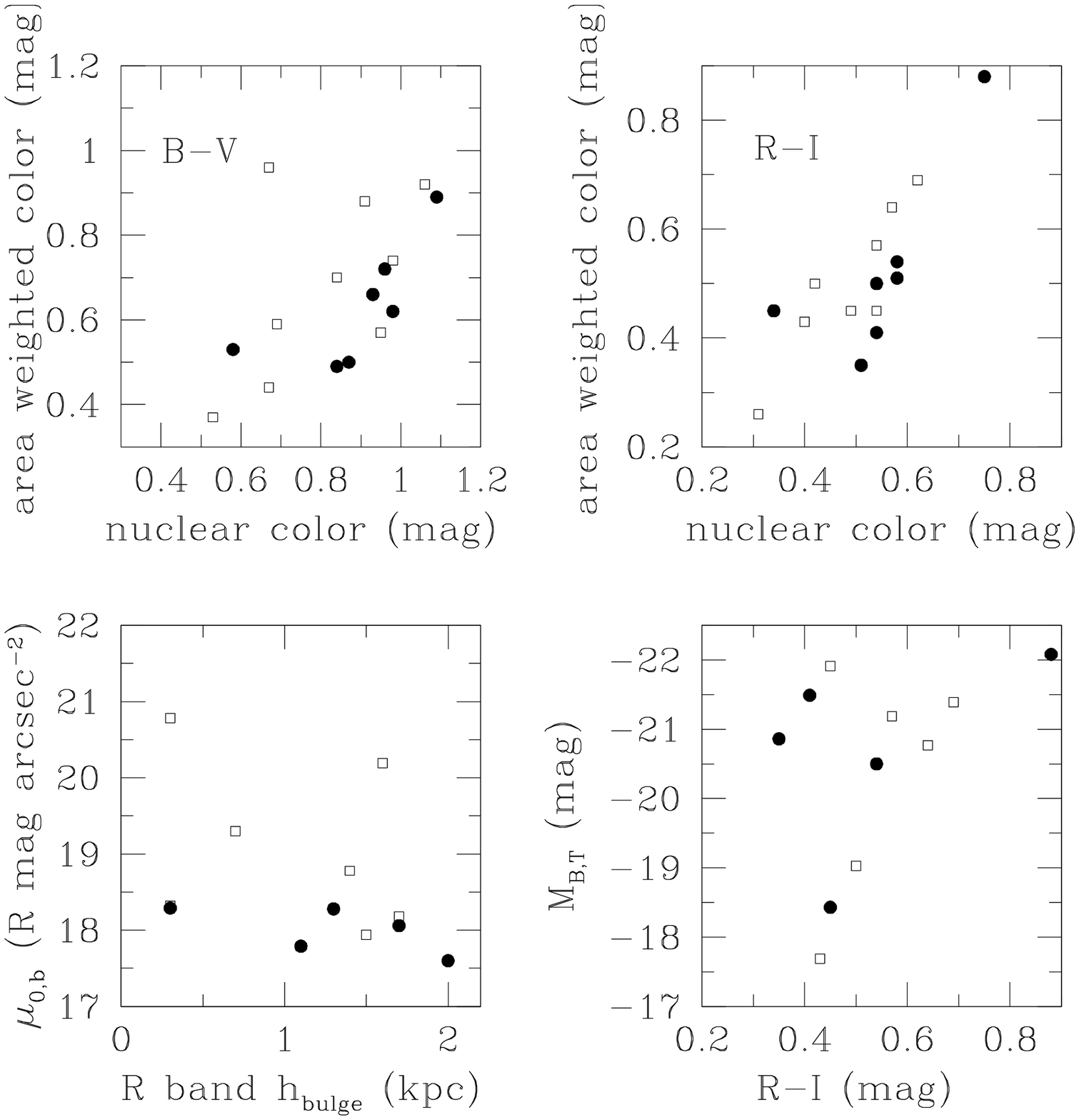}}
\caption[]{Top-left: \emph{B-V} area weighted color versus \emph{B-V}
nuclear color. Top-right: \emph{R-I} area weighted color versus
\emph{R-I} nuclear color. Bottom-left: \emph{R} band central bulge
surface brightness versus \emph{R} band bulge scale
length. Bottom-right: Absolute \emph{B} magnitude versus \emph{R-I}
area weighted color. Open squares = LSB galaxies with round bulges. Filled dots = LSB galaxies with bars and rings. No
systematic differences between the two types of LSB galaxies show up
in these properties.} 
\label{types}
\end{figure}

Figure~\ref{types} shows the relation between various galaxy
properties as derived from the photometry for both types of LSB
galaxies present in our sample. In the two top panels both types
follow roughly the expected trend of having redder nuclear colors than
area weighted colors. In the bottom-left panel it is seen that the
bulges do not follow the same trend as the disks; there are galaxies
with large bulge scale lengths and high bulge central surface
brightnesses. The galaxies with bars appear to have relatively high
bulge central surface brightnesses, while the ones with round bulges
show a large(r) spread. The bottom-right panel shows the division in high
and lower luminosity galaxies, but both types are mixed. 

On the basis of disk or bulge colors, scale lengths, surface
brightnesses or absolute magnitudes no clear distinction can be made between
the two types. 

The correlation between bulge and disk scale lengths suggests
that the formation of these two components is closely coupled. The restricted range
of the ratio between bulge and disk scale lengths of $\sim 0.1$
is used as an argument for secular evolution models (Courteau et al. \cite{courteau}). In this picture bulges
form from small dynamical instabilities of disks which induce
formation of bars. Secular accretion or satellite accretion in the
center in turn can dissolve the bar into a spheroidal component
(Norman et al. \cite{norman}). If this scenario is correct then the bulges
are relatively young.

However, the observed color gradients suggest that at larger radii the emitted
light is dominated by a relatively young population of stars. This and
the fact that the bulges are redder than the disks favors a scenario
in which the bulge forms before the disk. It is hard to pin down the
exact causes of the color gradients without metallicity and extinction
information. Andredakis (\cite{andredakis2}) has shown that a correlation between bulge and
disk scale lengths also quite naturally arises if the bulge were to be formed
before the disk. If this scenario is correct
then the bulges are relatively old. 

By examining the bottom panels of Figs.~\ref{profiles} we see that the \emph{R-I} gradient is in general less steep than the \emph{B-V}
and \emph{B-R} gradients and the \emph{R-I} profile is in most cases fairly constant. Comparison
of bulge and disk colors shows that the color difference is smallest
in \emph{R-I}. This is reflected in the fact that the differences in
B/D ratios are also smallest between \emph{R} and
\emph{I} band for most of the galaxies in our sample. This could
indicate that the old stellar populations in the bulge and disk are of
approximately the
same age. But without knowledge of metallicity, extinction, initial mass function, and/or star
formation history it is hard to make any definite statements about the
difference in age between bulge and disk. The effects that differences in these properties have on
the colors of disk galaxies are extensively discussed in van den Hoek
et al. (A\&A submitted). Peletier \& Balcells
(\cite{peletier_balcells}) conclude from their study of optical and
near-infrared colors of 30 galaxies of
types S0 to Sbc that the inner disks are only slightly younger than
bulges. The difference in age generally lies somewhere between 0 and 3
Gyrs.

\clearpage

\begin{figure}[!t]
\resizebox{\hsize}{!}{\includegraphics{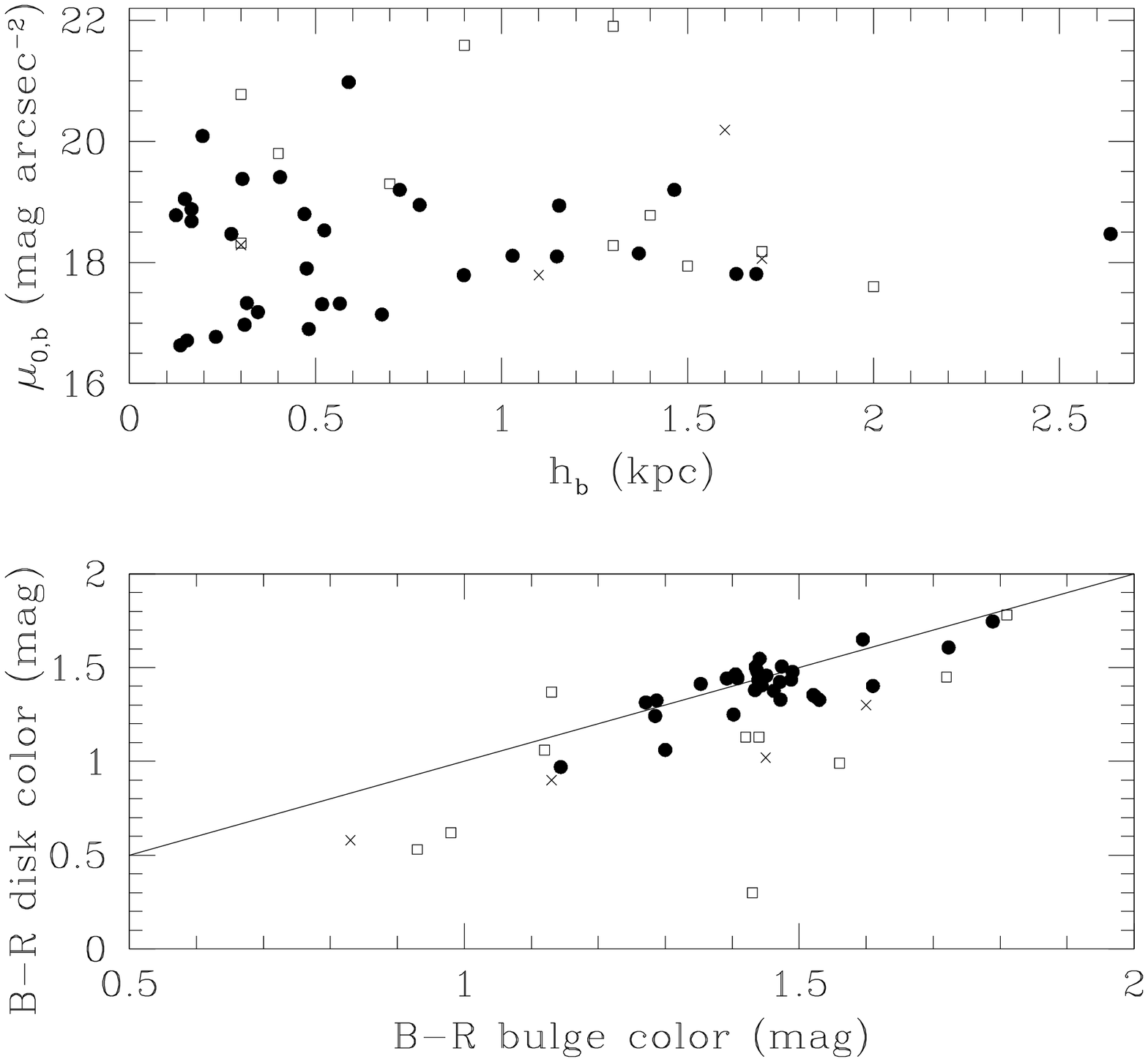}}
\caption[]{Upper panel: Distribution of \emph{R} band bulge central
surface brightness with bulge scale length. Filled circles = Sb,Sc
type HSB galaxies (Andredakis
1997) . Diagonal
crosses = Sb,Sc type LSB galaxies in our sample (Table 1). Open
squares = other types of LSB
galaxies in our sample. Note that all galaxies have large scatter in bulge central
surface brightness and that there are galaxies with long bulge scale
lengths and relatively high surface brightnesses. Lower panel:
\emph{B-R} disk color versus \emph{B-R} bulge color. Filled circles = Sb,Sc
type HSB galaxies (Peletier \& Balcells 1996). The other symbols have 
the same meaning as in the upper panel. The diagonal line indicates
the locus where both colors are equal.}
\label{bulgepar}
\end{figure}

A natural question that arises is whether the bulges of LSB galaxies
are different than those of HSB galaxies. To explore this question we compared
the bulge parameters, central surface brightness (\emph{R} band) and scale length, with
a sample of 33 Sb,Sc type HSB galaxies (Andredakis 1997, see upper
panel of Fig.~\ref{bulgepar}). The bulge
colors were compared with another sample of 30 S0--Sbc type galaxies (Peletier \& Balcells
\cite{peletier_balcells}, see bottom panel of Fig.~\ref{bulgepar}). Andredakis' sample
uses \emph{r} band photometry (Thuan \& Gunn \cite{thuan_gunn}), but
that has no effect on our conclusions. Peletier \& Balcells take the
bulge color at 0.5$\times R_{\rm eff}$ or at 5\arcsec, whichever is
larger and the disk color at 2 \emph{K} band scale lengths. We take
the nuclear color as bulge color and the color at 2 \emph{R} band disk
scale lengths as disk color. No distinction can be made between
bulges of HSB and LSB galaxies on basis of their structural
parameters. The distribution of bulge colors of the LSB galaxies has a
larger spread, but on average the bulges of HSB and LSB galaxies have similar colors. LSB disks are bluer making the differences between bulge and disk colors  
larger for LSB galaxies. 

We note that all our conclusions are based on a small
sample of LSB galaxies and it would clearly help if data for more bulge LSB
galaxies would become available.

\section{Summary}

We have presented \emph{BVRI} imaging of a sample of LSB galaxies with bulges. We find that these
objects are well described as exponential disks with
exponential bulges. The disk
and bulge scale lengths of the galaxies have no preferred value. The central surface brightness does not depend
on scale length or color, excluding fading scenarios. Despite the
median central surface brightness of $\mu_{0,\rm d}$ = 23.04
mag arcsec$^{-2}$ the median scale length ($h_{\rm d}$ = 12.6 kpc) is very
large. The largest galaxies have sizes and luminosities comparable to
giant LSB galaxies.

By decomposing the luminosity
profiles using double exponentials we find that the disk scale lengths
are correlated with the
bulge scale lengths. This suggest a coupling in the formation of bulge and disk.

The bulge dominated LSB galaxies are observed to be redder than disk
dominated LSB galaxies. On average the
colors are still \emph{bluer} than the colors of typical Freeman
galaxies. The B/D ratios are observed to increase towards the
longer wavelengths indicating that the bulges of LSB galaxies are redder than their
disks.

The ratios between bulge and disk scale lengths scatter around $\sim
0.1$ and the increase in
B/D ratios towards longer wavelengths is consistent with
observations of HSB spirals (dJ95). From the
images shown in the appendix it is apparent that LSB galaxies exist in a wide range of morphological
types. There seem to be roughly two types of bulge LSB galaxies in
our sample. There are the ones with normal (round) bulges and the ones
that look like they are undergoing some heavy evolution with blue
rings and bars. But besides appearance no
systematic differences show up in the rest of their investigated properties.

It remains unclear whether the bulge formed before or after the
disk. More observations are needed to resolve this issue. We did not
find any systematic differences in structural parameters or colors
of the
bulges between LSB and HSB galaxies.

To unravel the evolutionary history of bulge dominated LSB galaxies we
need additional information on \ion{H}{i} content and distribution, and
chemical abundance information. Our sample has extended the fact
that LSB galaxies cover a wide range in color, luminosity, size and
morphology. The bulge dominated LSB galaxies fit in with the general
trends defined by their HSB counterparts. 

\begin{acknowledgements}

We thank Adwin Boogert, Jos de Bruyne and Lodewijk Bonebakker for
obtaining the data.

We thank Roelof de Jong for making his data of a large
sample of spirals available to us.

This research has made use of the NASA/IPAC Extragalactic Database 
(NED) which is operated by the Jet Propulsion Laboratory,
California Institute of Technology, under contract with the National
Aeronautics and Space Administration.

\end{acknowledgements}

\listofobjects

\appendix
\section{R band images of sample galaxies}
\begin{figure*}[!t]
\resizebox{\hsize}{!}{\includegraphics{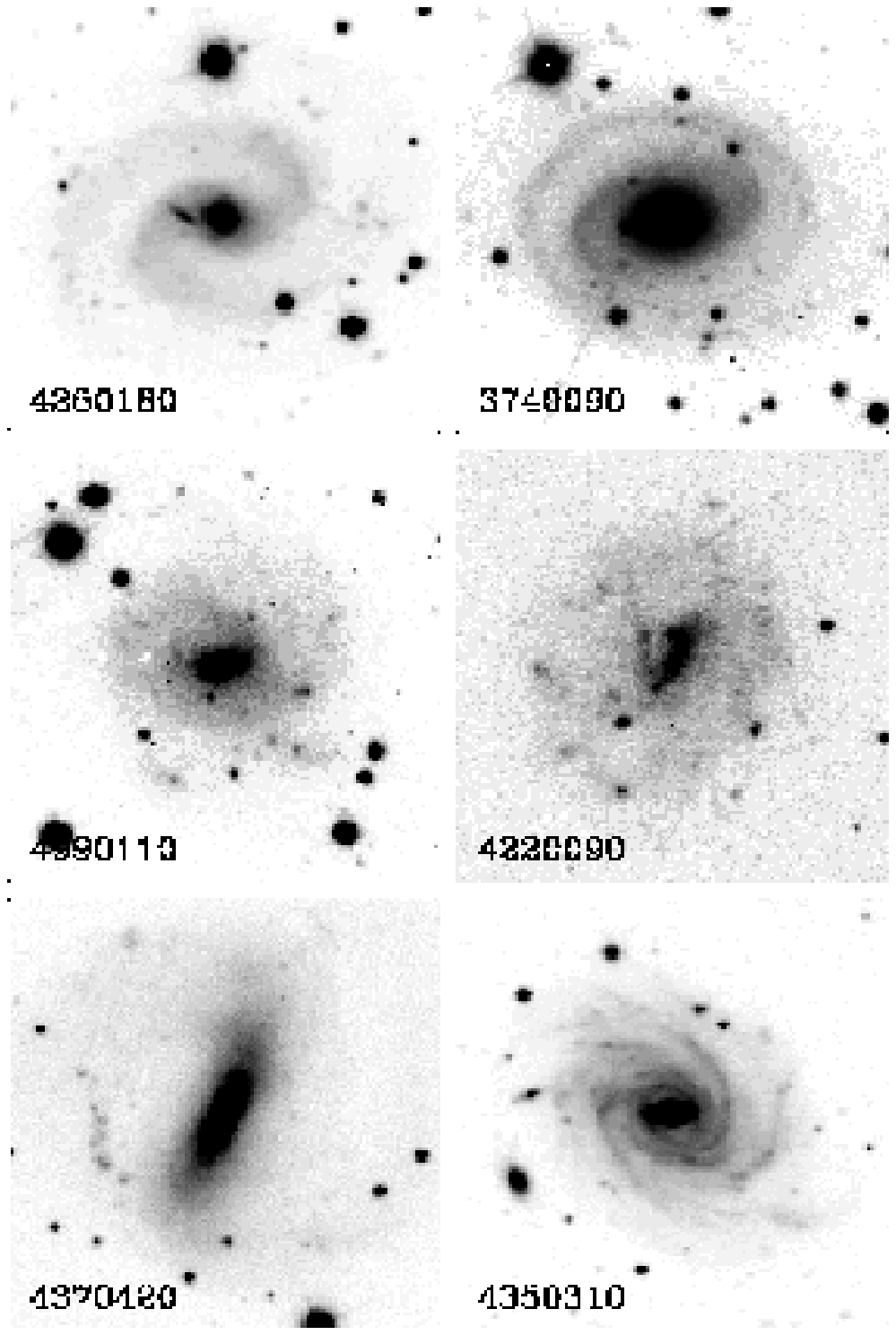}}
\caption[]{R band images of the sample of LSB galaxies. Top-left:
ESO-LV 4250180; top-right: ESO-LV 3740090; middle-left: ESO-LV 4990110;
middle-right: ESO-LV 4220090; bottom-left: ESO-LV 4370420;
bottom-right: ESO-LV 4350310. The image of ESO-LV 3740090 is displayed
using exponential intensity scale. Image sizes are
1.48 $\times$ 1.48 arcmin$^2$.} 
\end{figure*}

\begin{figure*}[!t]
\resizebox{\hsize}{!}{\includegraphics{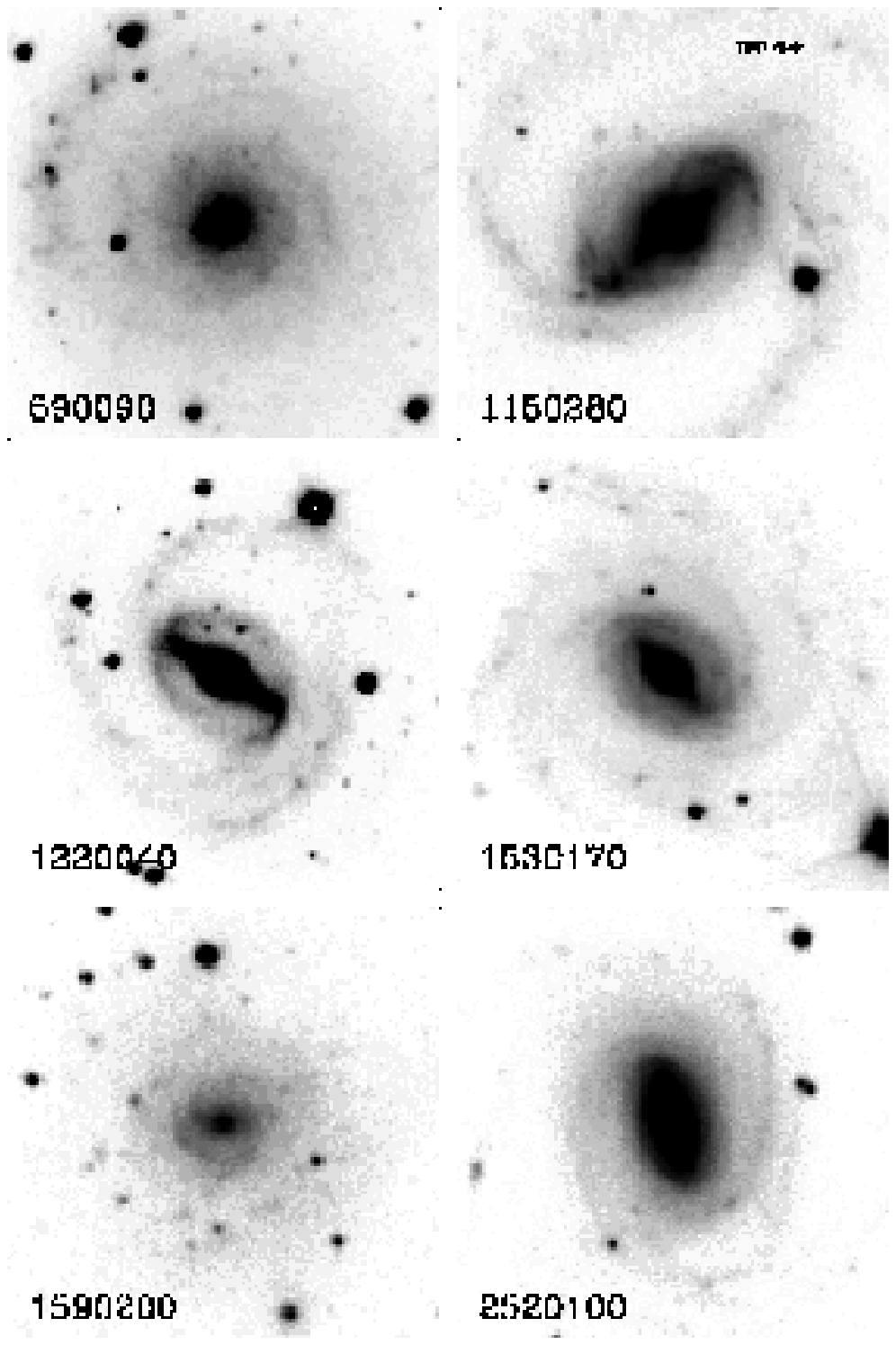}}
\caption[]{R band images of the sample of LSB galaxies. Top-left:
ESO-LV 590090; top-right: HSB galaxy ESO-LV 1150280; middle-left:
ESO-LV 1220040;
middle-right: ESO-LV 1530170; bottom-left: ESO-LV 1590200;
bottom-right: ESO-LV 2520100. The images of ESO-LV 1150280, 1530170, 1590200 and
2520100 are displayed using exponential intensity scale. Image sizes are
1.48 $\times$ 1.48 arcmin$^2$.} 
\end{figure*}

\begin{figure*}[!t]
\resizebox{\hsize}{!}{\includegraphics{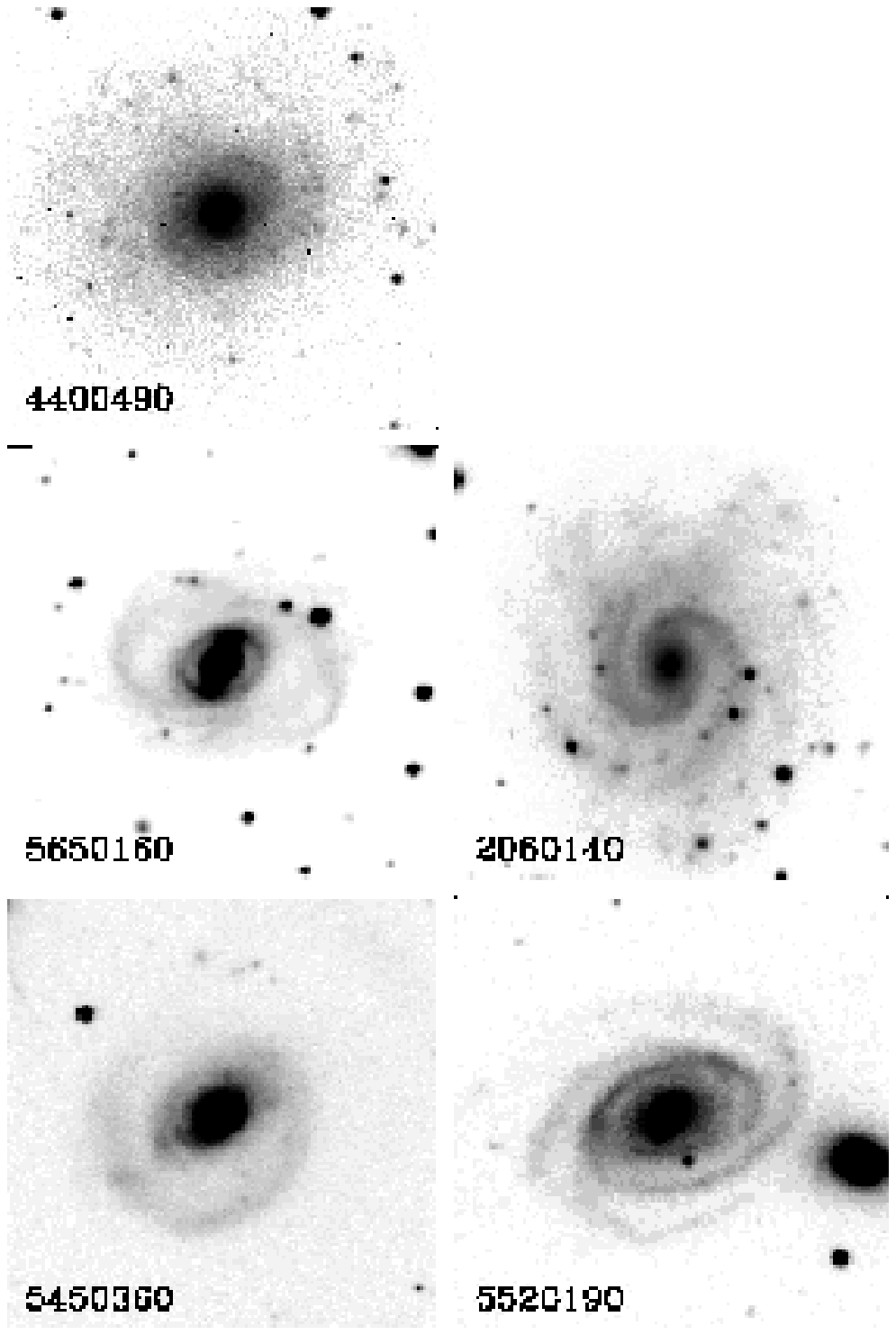}}
\caption[]{R band images of the sample of LSB galaxies. Top-left:
ESO-LV 4400490; middle-left: ESO-LV 5650160;
middle-right: ESO-LV 2060140; bottom-left: ESO-LV 5450360;
bottom-right: ESO-LV 5520190. The images of ESO-LV 4400490 and 2060140 are
displayed using exponential intensity scale. Image sizes are
1.48 $\times$ 1.48 arcmin$^2$.} 
\end{figure*}

\begin{figure*}[!t]
\resizebox{\hsize}{!}{\includegraphics{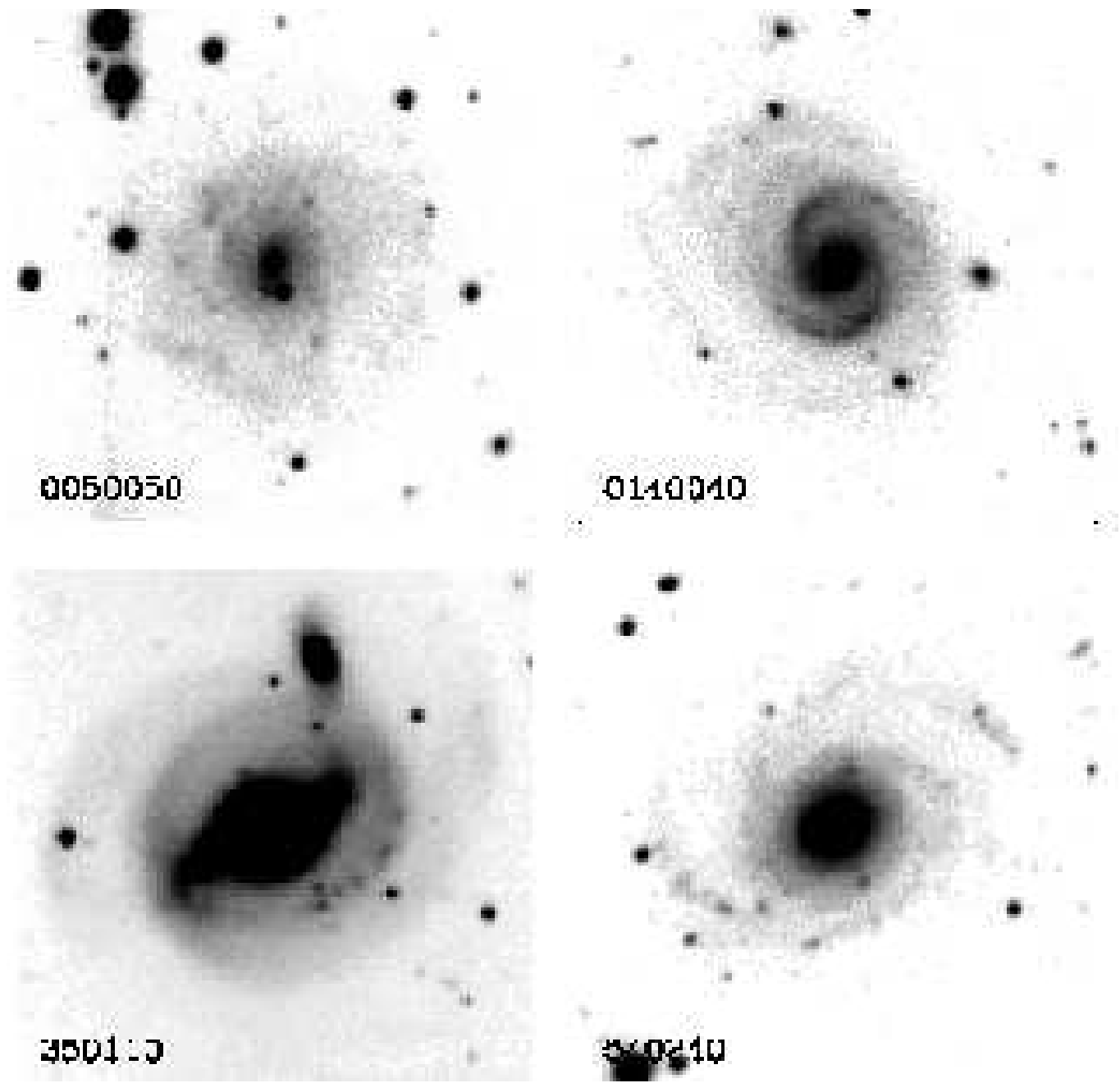}}
\caption[]{R band images of the sample of LSB galaxies. Top-left:
ESO-LV 0050050; top-right: ESO-LV 0140040; bottom-left: ESO-LV 350110;
bottom-right: ESO-LV 540240. The images of ESO-LV 0050050, 0140040 and 540240
are displayed using exponential intensity scale. Image sizes are
1.48 $\times$ 1.48 arcmin$^2$.} 
\end{figure*}
\end{document}